# Linear stability analysis of particle-laden Couette-Poiseuille flows: effect of porous walls

**Ananthapadmanabhan Ramesh[1], Abbas Moradi Bilondi[1], Mohammadreza Mahmoudian[1] and , Parisa Mirbod[1]†**

[1]Department of Mechanical and Industrial Engineering, 842 W. Taylor Street, University of Illinois at Chicago, Chicago, IL 60607, USA

The current study presents a three-dimensional linear stability analysis of particle-laden Couette-Poiseuille flow suspended in a Newtonian fluid between two parallel plates, with the lower plate coated by a porous medium. The influence of suspended particles is examined using a two-domain formulation in which particles are confined to the fluid layer and do not penetrate the porous substrate. The particle-laden suspension is modeled using the dusty-gas framework, while the flow within the porous layer is described by the volume-averaged Navier-Stokes (VANS) equations. In particle-laden flows over impermeable walls, particle inertia may either stabilize or destabilize the flow depending on the governing parameters. In contrast, the presence of a porous layer introduces an additional permeability-dependent destabilizing mechanism that fundamentally modifies these classical trends. Consequently, particle loading can reduce the critical Reynolds number at sufficiently high permeability, even in parameter regimes where particles stabilize the corresponding rigid-wall flow. The coupled formulation also introduces additional disturbance branches associated with fluid-particle coupling near the permeable interface. Although these modes remain stable throughout the parameter space investigated, they modify the eigenspectrum and influence the dominant instability through altered coupling pathways. Furthermore, unlike impermeable-wall Couette-Poiseuille flow, where increasing the Couette component generally stabilizes the flow, the porous-wall configuration exhibits a monotonic decrease in the critical Reynolds number over the range examined. These results demonstrate that porous boundaries can fundamentally alter established stability behavior in particle-laden shear flows through permeability-dependent coupling between the suspension and the porous substrate.

**Key words:** Particle-fluid flow, dusty gas model, porous surfaces, volume-average Navier-Stokes equation

## 1. Introduction

The study of flow dynamics in particle-laden fluids over porous surfaces has attracted substantial attention due to its relevance in a range of natural, biological, and industrial systems. In biological contexts, such flows play a critical role in understanding transport mechanisms in blood vessels, the gastrointestinal tract, and the renal system (Majdalani *et al.* 2002; Chang *et al.* 1989; Mirbod *et al.* 2017; Haffner & Mirbod 2020). Similarly, in industrial applications, porous media and particle-laden flows are prevalent in filtration systems, oil recovery processes, and chemical reactors, making it essential to comprehensively investigate their stability characteristics (Blest *et al.* 1999; Allen 1984; Perazzo *et al.* 2018; Ewing & Weekes 1998). While biological systems motivate the broader relevance of particle-porous interactions, the present formulation is restricted to dilute inertial suspensions and does not attempt to quantitatively model complex suspensions such as blood.

† Email address for correspondence: pmirbod@uic.edu

Numerous studies have examined the stability of plane Poiseuille flow (PF) in fluid-porous media systems to understand the interaction between fluid dynamics and porous structures. Chang *et al.* (2006) conducted a linear stability analysis of PF overlying a porous substrate, considering perturbations with random wavenumbers. Employing the Beavers and Joseph slip condition (Beavers & Joseph 1967*a*), they identified two distinct instability modes: a fluid-layer mode and a porous-layer mode. The study emphasized that the depth ratio and the transition layer thickness critically influence flow stability. Expanding on these findings, Liu *et al.* (2008*a*) employed the Darcy-Brinkman model to provide a more refined representation of the porous medium. They observed two unstable modes but noted the absence of an odd-fluid-layer mode due to velocity continuity at the fluid-porous interface, leading to symmetry in both the base and perturbed states. In contrast, Hill & Straughan (2008*a*) introduced a three-layer framework that included a Darcy-Brinkman layer sandwiched between a Darcy layer and a fluid layer. They reported that the neutral stability boundaries lost their bimodal characteristics under certain parameter regimes.

The stability of Couette-Poiseuille flow (CPF) in smooth channels has also been extensively examined due to its practical significance in lubrication systems, cylindrical reactors, and viscometers (Joseph & Saut 1990; Joo & Shaqfeh 1994; Snoeijer & der Weele 2014). Potter (1966) analyzed the linear stability of CPF between two infinite smooth parallel plates using the Orr-Sommerfeld equation, demonstrating that the addition of Couette flow generally stabilizes the flow, except when the upper wall velocity ($W$) is approximately 0.2-0.4 times the maximum Poiseuille velocity ($U$). Furthermore, Hains (1967) extended the analysis to the entire range of wall velocities, noting that at higher $W$, the critical Reynolds number increased indefinitely, irrespective of whether the wall moved upstream or downstream.

More recent studies have focused on the influence of porous media on CPF. Chang *et al.* (2017) was the first to explore the linear stability of CPF over porous substrates, revealing that at small depth ratios, Couette flow can destabilize the system, producing a tri-modal structure in the neutral stability curves. However, at higher upper wall velocities, the dominant instability mode shifted from a fluid-layer mode to a porous-layer mode. In a subsequent study, Hooshyar *et al.* (2022) coupled the Brinkman and Navier-Stokes equations to assess the stability of CPF over a porous layer. They introduced the concept of a cutoff velocity, a critical wall velocity at which the imposed Couette component no longer influences the stability of Poiseuille flow over a porous boundary, with its value shown to decrease as the permeability of the porous medium is reduced or as the fluid layer thickness increases. Despite these advances, limited attention has been directed towards particle-laden flows over porous media, particularly in configurations involving both a pressure gradient and Couette flow. Understanding such interactions is crucial for environmental and industrial applications where multiphase flows interact with complex substrates.

Several studies have investigated the effect of porous media on flow stability in planar configurations. Silin *et al.* (2011) analyzed the flow instability in a configuration in which a porous medium partially obstructs the flow, allowing for penetration. Their combined theoretical and experimental study showed strong agreement between the two approaches, underscoring the significant impact of the depth ratio on stability. Similarly, Samanta (2017) explored the influence of slip length and depth ratio, showing that the stability characteristics differed significantly from those observed in cases where a porous slab replaced a slippery upper wall. Tilton & Cortelezzi (2006) conducted a three-dimensional stability analysis of laminar flow confined between two porous slabs using volume-averaged Navier-Stokes (VANS) equations. They demonstrated that even marginal increases in wall permeability could substantially amplify instability, underscoring the pronounced effect of porous walls on flow stability. Parallel to these studies, recent research has focused on particle-laden flows, revealing that inhomogeneous particle distributions can significantly alter flow stability. Rouquier *et al.* (2019) employed a continuum model of particle-laden pipe flow coupled via Stokes drag and demonstrated that preferential particle localization could introduce linear instability, providing an alternative route to turbulence. These findings were corroborated by experimental studies by Suga *et al.* (2018), who observed a transition from laminar to turbulent flow in particle-laden channels. In particular, Rouquier *et al.* (2019) also demonstrate that the effect of particle inertia on stability is generally non monotonic, with intermediate relaxation times producing the strongest destabilization. The present work does not seek to establish this behavior as new rather, it examines how the introduction of a porous boundary alters these established particle-induced stability trends.

Modeling particle-laden flows typically requires accounting for particle-fluid coupling, and several numerical methods have been developed to simulate the dynamics of suspended particles. Review articles by Maxey (2017) and Balachandar & Eaton (2010) provide comprehensive insights into the numerical treatment of multiphase flows at low to moderate Reynolds numbers. In such systems, dilute suspensions are often analyzed using the dusty gas model, where particles are treated as a secondary fluid phase interacting with the primary flow (Saffman 1962). Expanding on this methodology, Mirbod *et al.* (2024) modeled dust-fluid interactions via Stokes drag, incorporating the effects of particle concentration and relaxation time. Klinkenberg *et al.* (2011) also conducted a non-modal instability analysis of dilute suspensions in channel flows, showing that flow stabilization depends strongly on the Stokes number, with higher values amplifying particle velocity perturbations. Boronin (2012) demonstrated that in channel flows with inhomogeneous particle distributions, maximum disturbance energy growth occurs when dust layers are located midway between the channel centerline and walls. Furthermore, Boronin & Osiptsov (2014) analyzed non-modal instability in laminar flows with spatially varying particle concentrations, demonstrating that particle distribution significantly influences flow stabilization in plates and channels. Beyond Stokes drag, additional fluid-particle interactions, including Basset history, buoyancy, and Saffman lift forces, become relevant in flows where the fluid-particle density ratio approaches unity (Maxey & Riley 1983; Saffman 1995; Boronin & Osiptsov 2008; Klinkenberg *et al.* 2014).

Despite advances in understanding flow stability in porous media and particle-laden flow systems separately, the interplay between these mechanisms under combined pressure-driven and shear-driven flows over porous substrates remains largely unexamined. This study addresses this gap by systematically investigating how suspended particles interact with porous boundaries under Couette-Poiseuille flow, uncovering new flow physics and instabilities driven by the coupling between flow structures, particle dynamics, and porous media properties. By varying permeability, particle parameters (including particle relaxation time and mass fraction), and the Couette component, while keeping the porous layer thickness and porous media porosity constant, we provide a comprehensive framework for predicting flow stability in complex multiphase systems, with implications for both fundamental understanding and industrial applications. The stability behavior observed here differs from that of particle-laden Couette–Poiseuille flow over impermeable walls, in that wall permeability modifies the known particle-induced stability trends and introduces additional coupling pathways between particle inertia and wall modes.

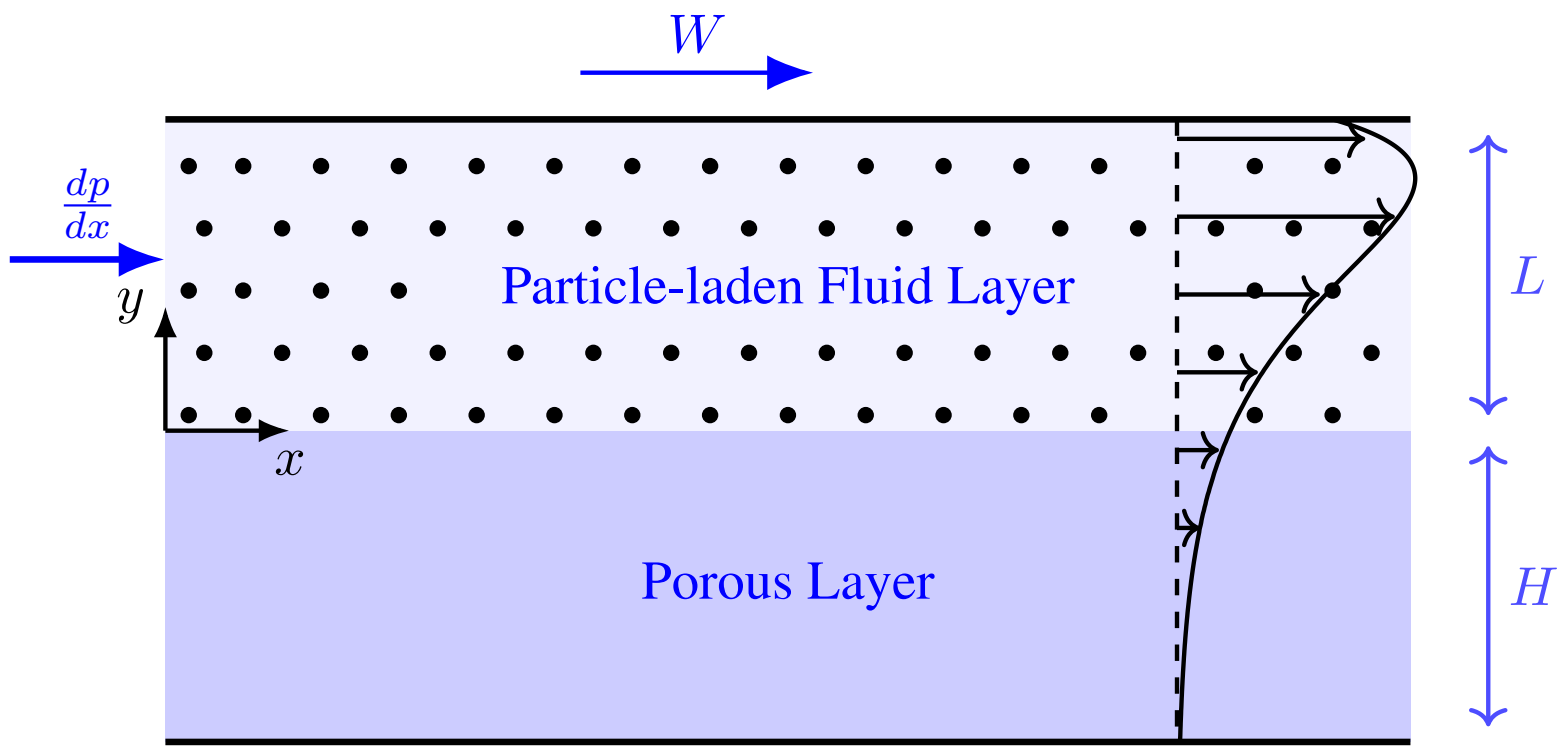

Figure 1: Schematic representation of the computational domain and coordinate system. The figure qualitatively illustrates velocity damping into the finite-thickness porous layer due to Darcy drag, Brinkman viscous diffusion near the interface, and the no-slip condition at the fixed bottom boundary (y = -H).

The remainder of this paper is organized as follows. Section 2 outlines the problem formulation, including the governing equations and boundary conditions (Section 2.1), the base (steady-state) flow configuration (Section 2.2), and the linear stability analysis methodology (Section 2.3). In Section 3, we present and discuss the results of the stability analysis, emphasizing the effects of permeability, Couette shear, and particle concentration on flow behavior. Finally, Section 4 summarizes the key findings and suggests directions for future research.

## 2. Problem formulation

We investigate a three-dimensional (3D) linear stability of a viscous, incompressible Newtonian fluid containing uniformly distributed particles in the free flow region of a channel of height L, with the lower portion occupied by a homogeneous, isotropic porous layer of thickness H, as shown in figure 1. The flow is driven in the streamwise direction by a uniform pressure gradient $G$, and by shear due to the constant velocity $W$ of the impermeable upper wall. The interface between the fluid and the porous medium lies at $y = 0$, with the bottom plate remaining fixed. The same fluid permeates both regions, and its density, viscosity, and surface tension are taken to be constant.

The interaction between fluid and particles is modeled using the dusty gas model introduced by Saffman (1962). This model is valid for dilute suspensions of small, rigid, spherical particles, whose diameter is much smaller than the characteristic length scale of the flow. It assumes that the dominant interaction between the particles and the fluid arises from Stokes drag, neglecting finite-size effects and particle-particle interactions. In the present study, we consider inertial, heavy particles whose density is significantly greater than that of the fluid and for which the drag force dominates. In many practical particulate flows, particularly at moderate to high concentrations, particle-particle interactions, collisions, hydrodynamic interactions, and particle-phase stresses can significantly influence the dynamics and have been extensively studied (Elghobashi 1994; Balachandar & Eaton 2010; Maxey 2017; Rosti *et al.* 2021; Demou *et al.* 2022; Mirbod *et al.* 2023). However for the present work, the dusty gas model is deliberately chosen since it is well established and appropriate for the dilute two-way coupled regime targeted in this study: volumetrically dilute suspensions (low particle volume fraction $\phi_{bulk} \ll 1$) with significant mass loading due to high particle to fluid density ratio, where Stokes drag dominates momentum exchange and particle-particle effects are negligible to leading order. This regime is relevant to many gas-solid applications, such as aerosols, dust transport, and dilute filtration. The present formulation is therefore not intended to quantitatively describe near-neutrally buoyant or dense suspensions, such as blood, where additional forces and finite-size or deformability effects become important. While caustic formation can limit the dusty gas continuum description to $St < 1$ in turbulent flows (Wilkinson & Mehlig 2005), the present linear stability analysis of a laminar base flow with uniform particle distribution preserves single-valued velocity fields. The model remains valid for arbitrary St in this regime, as confirmed in related linear stability studies (Klinkenberg *et al.* 2011; Sozza *et al.* 2022). More advanced models are needed for denser regimes (four-way coupling), but introducing closures for particle stresses, collisions, or additional forces (Basset history, lift) would add substantial complexity to the two-domain linear stability analysis (dusty gas model coupled with VANS with interface conditions) and obscure the core mechanisms under investigation: the interplay between particle inertia (relaxation time $S$), concentration (mass fraction $f$), and porous-wall permeability in modulating stability. In addition, the dusty gas formulation has been widely adopted and validated for linear stability in similar dilute parallel shear flows, successfully capturing particle stabilization due to inertia (Saffman 1962; Klinkenberg *et al.* 2011; Boronin & Osiptsov 2014), and has recently been extended to porous boundaries (Mirbod *et al.* 2024).

### 2.1. *Governing equations and boundary conditions*

The governing mass and momentum equations for modeling fluid flow with particles are given by the dusty gas model (Saffman 1962) as

$$\nabla \cdot \boldsymbol{u} = 0, \tag{2.1}$$

$$\rho\left(\frac{\partial \boldsymbol{u}}{\partial t} + \boldsymbol{u} \cdot \nabla \boldsymbol{u}\right) = -\nabla p + \mu \nabla^2 \boldsymbol{u} - n\boldsymbol{F}_{St}, \tag{2.2}$$

$$\frac{\partial n}{\partial t} + \nabla \cdot (n\boldsymbol{u_p}) = 0, \tag{2.3}$$

$$\frac{4\pi a^3}{3}\rho_p\left(\frac{\partial \boldsymbol{u_p}}{\partial t} + \boldsymbol{u_p} \cdot \nabla \boldsymbol{u_p}\right) = \boldsymbol{F}_{St}. \tag{2.4}$$

Where $\boldsymbol{u}$, $\boldsymbol{u_p}$, $p$, $\rho$, $\rho_p$, $\mu$ and $n$, are the velocity of the pure fluid, velocity of particles, pressure, density of the fluid, density of particles, dynamic viscosity and particle number density. The coupling between the pure fluid and the particle is given by the Stokes drag force and is indicated as $\boldsymbol{F}_{St} = 6\pi\mu a\,(\boldsymbol{u} - \boldsymbol{u_p})$, where $a$ is the radius of the spherical particle. In the dusty gas model, the carrier fluid is treated as incompressible (equation 2.1), while the particle number density $n$ is allowed to vary (equation 2.3). This apparent asymmetry is a standard asymptotic approximation valid for volumetrically dilute suspensions ($\phi_{bulk} \ll 1$), where particles occupy negligible volume and do not significantly displace the fluid. Local changes in particle concentration, therefore, do not require compensating variations in fluid density or volume fraction at the leading order retained in the model. Instead, concentration fluctuations primarily influence mixture inertia and interphase momentum exchange via Stokes drag. For non-dilute regimes where particle volume effects become significant, more general two-fluid models enforcing mixture incompressibility would be required.

Solving the Navier-Stokes and continuity equations within a porous medium presents significant challenges due to the broad range of length scales involved, spanning from microscopic pore diameters to the macroscopic characteristic thickness of the porous layer ($H$). To address this complexity, a volume-averaging technique is employed to focus on the macroscopic behavior of the flow. In the present study, the particle-laden flow in the channel is coupled with the volume-averaged Navier-Stokes (VANS) equations to describe the flow dynamics within the porous medium (Beavers & Joseph 1967*b*). While several methods have been proposed for deriving flow equations in porous media, and earlier studies have highlighted inconsistencies among these formulations, the volume-averaging technique has emerged as a reliable framework for consistently coupling channel flows with adjacent porous layers (Tilton & Cortelezzi 2008; Rosti *et al.* 2021; Hooshyar *et al.* 2022; Mirbod *et al.* 2023, 2024). However, it is important to note that the VANS formulation becomes invalid in the thin heterogeneous transition region near the channel-porous interface due to abrupt structural variations. To address this, Ochoa-Tapia & Whitaker (1995, 1998) introduced momentum transfer conditions that incorporate modifications in the transition layer and apply a jump condition in the shear stress to account for discontinuities. Furthermore, the present analysis assumes negligible inertial effects within the porous medium, thereby simplifying the VANS equations by omitting the convective and Forchheimer terms. This assumption is justified for porous media with low permeability, where the interfacial velocity remains significantly smaller than the peak velocity in the adjacent channel. The governing mass and momentum conservation equations for the porous layer are then given as

$$\nabla \cdot \boldsymbol{u_m} = 0, \tag{2.5}$$

$$\frac{\rho}{\epsilon}\frac{\partial \boldsymbol{u}_m}{\partial t} = -\nabla p_m + \mu_e \nabla^2 \boldsymbol{u}_m - \frac{\mu}{\kappa}\boldsymbol{u}_m. \tag{2.6}$$

Here $\boldsymbol{u_m}$ and $p_m$ denote the Darcy-scale velocity and pressure fields, $\epsilon$ is the porosity, $\mu_e$ is the effective viscosity accounting for the slip at the fluid-porous interface (Wu & Mirbod 2019), and $\kappa$ is the permeability of the porous material. The porous medium parameters used in the present study are chosen based on experimentally measured and theoretically reported structural properties of representative porous materials, as summarized in Table 1. For a representative channel height $L$ on the order of millimeters to centimeters (typical of laboratory filtration or coating systems), this range corresponds approximately to intrinsic permeabilities spanning $10^{-7}$ to $10^{-10}$ m$^2$. At the fluid-porous interface ($y = 0$), we impose a zero flux condition for the particle phase where the normal particle flux vanishes, equivalent to no penetration of particles into the porous layer. This is implemented via the no penetration condition on the wall normal particle velocity perturbation and is consistent with the assumption that particles are either too large relative to the pore size or blocked by a filtration interface. Accordingly, no particle transport equations are solved within the porous domain. This modeling approach is consistent with prior studies that neglect particulate motion in porous media under strong size-exclusion or filtration conditions (Beavers & Joseph 1967*b*; Nield *et al.* 2006).

The boundary conditions for the coupled system are specified at the channel walls and at the fluid-porous interface. At the moving upper wall ($y = L$), a no-slip condition is imposed as

$$u = W; v = 0; w = 0, \tag{2.7}$$

where $u$, $v$, and $w$ are the streamwise, wall-normal, and spanwise velocity components, respectively, and $W$ denotes the constant upper wall velocity. At the fixed lower boundary of the porous region ($y = -H$), no-slip conditions are applied to the averaged velocity field within the porous medium, given by

$$u_m = v_m = w_m = 0, \tag{2.8}$$

where $u_m$, $v_m$, and $w_m$ represent the streamwise, wall-normal, and spanwise components of the Darcy-averaged velocity inside the porous layer.

| *Material* | $\epsilon$ | $\alpha$ |
|---|---|---|
| Glass bead 1 | 0.38 | 34.678 |
| Foametal B | 0.78 | 50.443 |
| Foametal A | 0.58 | 101.639 |
| Aloxite | 0.58 | 393.749 |

Table 1: The experimental and theoretical values of structural properties of different porous materials (Samanta 2020)

In the current study, we assume velocity and pressure continuity at the suspension-porous interface, whereas there is a jump in the tangential stress, which depends upon the momentum transfer coefficient $\tau$ (Ochoa-Tapia & Whitaker 1995). Hence, the boundary condition at the fluid-porous interface ($y = 0$) is given as:

$$\begin{aligned}
u = u_m, \quad v = v_m, \quad w = w_m,\\
p_m - 2\mu_e \frac{\partial v_m}{\partial y} = p - 2\mu \frac{\partial v}{\partial y},\\
\mu_e \frac{\partial u_m}{\partial y} - \mu \frac{\partial u}{\partial y} = \frac{\tau\mu}{\sqrt{\kappa}} u_m,\\
\mu_e \frac{\partial w_m}{\partial y} - \mu \frac{\partial w}{\partial y} = \frac{\tau\mu}{\sqrt{\kappa}} w_m.
\end{aligned} \tag{2.9}$$

The coefficient $\tau$ quantifies the transfer of stress between the suspension fluid and the porous medium, and its value (zero, positive, or negative) depends on the structural characteristics of the porous material within the heterogeneous transition layer, as well as the interface. Alazmi & Vafai (2001) provides a comprehensive comparison of boundary conditions, while later studies investigated $\tau$ theoretically (Goyeau *et al.* 2003; Min & Kim 2005; Valdés-Parada *et al.* 2007, 2009) and, more recently, Carotenuto & Minale (2011) and Bagheri & Mirbod (2022) attempted its experimental measurement. In the present study, we set $\tau = 0$, indicating that the stress of the suspension fluid is completely transferred to the flow within the porous layer. The coupled dusty gas and porous medium formulation assumes a separation of length scales such that the characteristic pore size $l_p$ satisfies $l_p \ll a \ll L$, where $a$ is the particle radius and $L$ is the channel height. The first inequality ensures particle exclusion from the porous layer, while the second permits a continuum treatment of the particle phase within the free-flow region. This regime is representative of filtration flows and dust-laden channel flows over porous coatings, but is not intended to describe highly confined biological suspensions where particle sizes are comparable to the channel scale.

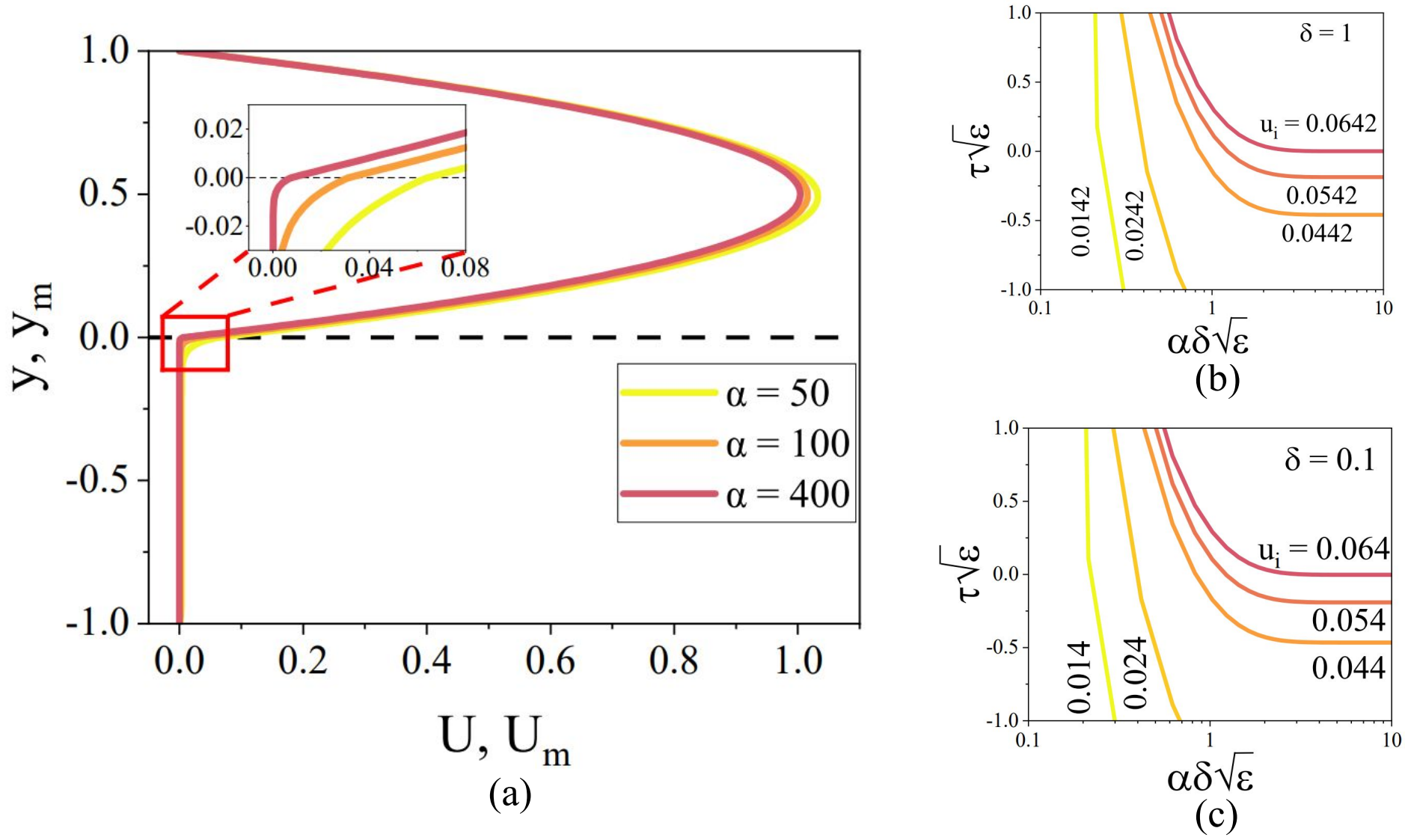

Figure 2: Base state results. (a) Velocity profiles of the base state obtained using equations 2.12 and 2.13 for different values of permeability parameter $\alpha$. Other parameters are $\epsilon$=0.6, $\tau$=0, $\delta$=1, and $W^*$=0. (b,c) Contours of the normalized slip velocity $u_i$ in the parameter space ($\alpha\delta\sqrt{\epsilon}$,$\tau\sqrt{\epsilon}$) for two different values of $\delta$ (b) $W^*$=0, $\delta$=1; (c) $W^*$=0, $\delta$=0.1.

### 2.2. *Base flow*

For CPF with suspended particles, the flow is driven by a moving plate and a constant pressure gradient, both acting in the streamwise direction. The flow maintains laminar characteristics because the wall velocity and pressure gradient are small, and variations are confined to the wall-normal direction ($u(y)$ and $u_m(y)$). We therefore consider a unidirectional, fully developed parallel base flow for which the wall normal and spanwise velocity components vanish ($v = v_m = 0$ and $w = w_m = 0$), which substantially simplifies the governing equations. The flow variables within the fluid layer are nondimensionalized as

$$\begin{aligned} u^* = u/U_{max} \quad , \quad v^* = v/U_{max} \quad , \quad w^* = w/U_{max}, \quad p^* = p/(\rho U_{max}^2), \\ t^* = t/(L/U_{max}) \quad , \quad x^* = x/L \quad , \quad y^* = y/L \quad , \quad z^* = z/L. \end{aligned} \tag{2.10}$$

Similarly, the flow variables within the porous layer are nondimensionalized as

$$\begin{aligned} u_m^* = u_m/U_{max} \quad , \quad v_m^* = v_m/U_{max} \quad , \quad w_m^* = w_m/U_{max}, \quad p_m^* = p_m/(\rho U_{max}^2), \\ t_m^* = t_m/(L/U_{max}) \quad , \quad x_m^* = x/L \quad , \quad y_m^* = y/L \quad , \quad z_m^* = z/L. \end{aligned} \tag{2.11}$$

Here, $U_{\max}$ denotes the maximum velocity of the pressure-driven (Poiseuille) component in the absence of wall motion. By solving the governing equations alongside the boundary conditions, the dimensionless base state velocity profiles within the fluid layer and porous layer are determined. For clarity of presentation, the superscript ($*$) used to denote non-dimensional variables has been dropped in the subsequent sections, except for $W$; all other quantities are treated as non-dimensional unless otherwise specified.

$$u(y) = u_p(y) = -\frac{1}{2}Gy^2 + (W^* + \frac{G}{2} - u_i)y + u_i, \tag{2.12}$$

$$u_m(y) = \frac{\left[(\tau G\alpha - \epsilon)\frac{G}{\alpha} + \epsilon\alpha(W^* + \frac{G}{2})\right]\sinh\left(A(1 + \frac{y_m}{\delta})\right)}{\alpha^2\sqrt{\epsilon}\left(\cosh A + \sqrt{\epsilon}(\frac{1}{\alpha} - \tau)\sinh A\right)}$$
$$+ \frac{2G\sinh\left(\frac{A}{2}(1 + \frac{y_m}{\delta})\right)\left\{\sqrt{\epsilon}\sinh\left(\frac{A}{2}(1 - \frac{y_m}{\delta})\right) + \epsilon(\frac{1}{\alpha} - \tau)\cosh\left(\frac{A}{2}(1 - \frac{y_m}{\delta})\right)\right\}}{\alpha^2\sqrt{\epsilon}\left(\cosh A + \sqrt{\epsilon}(\frac{1}{\alpha} - \tau)\sinh A\right)}. \quad (2.13)$$

The interface velocity $u_i$ between the fluid and the porous layer is given by

$$u_i(y) = \frac{\frac{G}{\alpha^2}(\cosh A - 1) + \frac{\sqrt{\epsilon}}{\alpha}(W^* + \frac{G}{2})\sinh A}{\left(\cosh A + \sqrt{\epsilon}(\frac{1}{\alpha} - \tau)\sinh A\right)}, \quad (2.14)$$

where $A = \alpha\delta\sqrt{\epsilon}$, $\alpha = \frac{L}{\sqrt{\kappa}}$ is the permeability parameter (with $\kappa$ the Darcy permeability), and $\delta = \frac{H}{L}$ denotes the depth ratio between the thickness of the porous layer $H$ and the height of the channel $L$. Since the permeability parameter $\alpha$ is inversely proportional to $\sqrt{\kappa}$, smaller values correspond to larger intrinsic permeability $\kappa$ of the porous medium. The imposed dimensionless pressure gradient is represented as $G = -\partial p/\partial x = 8$.

Equation 2.12 demonstrates that the base velocity field in the fluid layer arises from the superposition of a pressure-driven Poiseuille flow and a shear-driven Couette flow induced by the motion of the upper wall. Importantly, these base-state velocity fields (equations 2.12-2.14) are independent of the particle number density $n = N$, since the Stokes drag term vanishes in the particle momentum equation 2.4 for laterally uniform, steady flows. Throughout this study, the base-state particle number density is prescribed as a uniform constant $N_0$ (Mirbod *et al.* 2024). The hyperbolic form of equation 2.13 arises from balancing Brinkman viscous diffusion and Darcy drag in the finite thickness layer with no slip at y = -$\delta$. This yields a thin interfacial transition layer (thickness $\sim \sqrt{\kappa}$), a bulk near-uniform Darcy region, and a thin near-wall boundary layer consistent with bounded Darcy-Brinkman flows (Chang *et al.* 2017; Samanta 2020).

Figure 2a shows the base-state velocity profile of the particle-laden fluid-porous system for different permeability parameters ($\alpha$), while maintaining $W^* = 0$, $\tau = 0$, $\delta = 1$, and $\epsilon = 0.6$. As $\alpha$ increases, the velocity in the porous layer decreases significantly, approaching a plug-flow profile, indicating minimal flow penetration into the porous medium. In the limit of large $\alpha$, the porous layer effectively behaves as an impermeable wall, justifying the omission of inertial effects in the porous domain. The inset of figure 2a highlights that the interface velocity $u_i$ decreases with increasing the permeability parameter $\alpha$. From equation 2.14, $u_i$ is determined by three non-dimensional parameters: $A$ ($\alpha\delta\sqrt{\epsilon}$), $\sqrt{\epsilon}/\alpha$ and $\tau\sqrt{\epsilon}$. Figure 2b and 2c shows the effect of varying $\delta$ on $u_i$. For $\alpha\delta\sqrt{\epsilon} < 1$, the velocity is relatively insensitive to $\tau$. However, as $\alpha\delta\sqrt{\epsilon}$ increases, $u_i$ becomes more responsive to the momentum transfer coefficient, with the influence of $\delta$ becoming less pronounced. Figure 3a presents the effect of varying the dimensionless Couette component ($W^*$) on the base-state velocity profile. Increasing $W^*$ elevates the overall velocity within the fluid layer, promoting momentum diffusion into the porous medium. This effect is particularly evident in figures 3b and 3c, which depict the interface velocity $u_i$ for different $\delta$ values at $W^* = 1$. Consistent with the zero-Couette case, the interface velocity remains largely insensitive to $\tau$ when $\alpha\delta\sqrt{\epsilon} < 1$. However, for higher values of $\alpha\delta\sqrt{\epsilon}$, the sensitivity of $u_i$ to $\tau$ increases while its dependence on $\delta$ decreases. It should be noted that in the limit of very low porous permeability (i.e., $\alpha \rightarrow \infty$), the base velocity in the porous layer vanishes ($u_m \approx 0$), and the interface approaches a no-slip condition with $u_i = 0$. In this regime, the stability characteristics asymptotically approach those of particle-laden Couette-Poiseuille flow over a rigid impermeable wall.

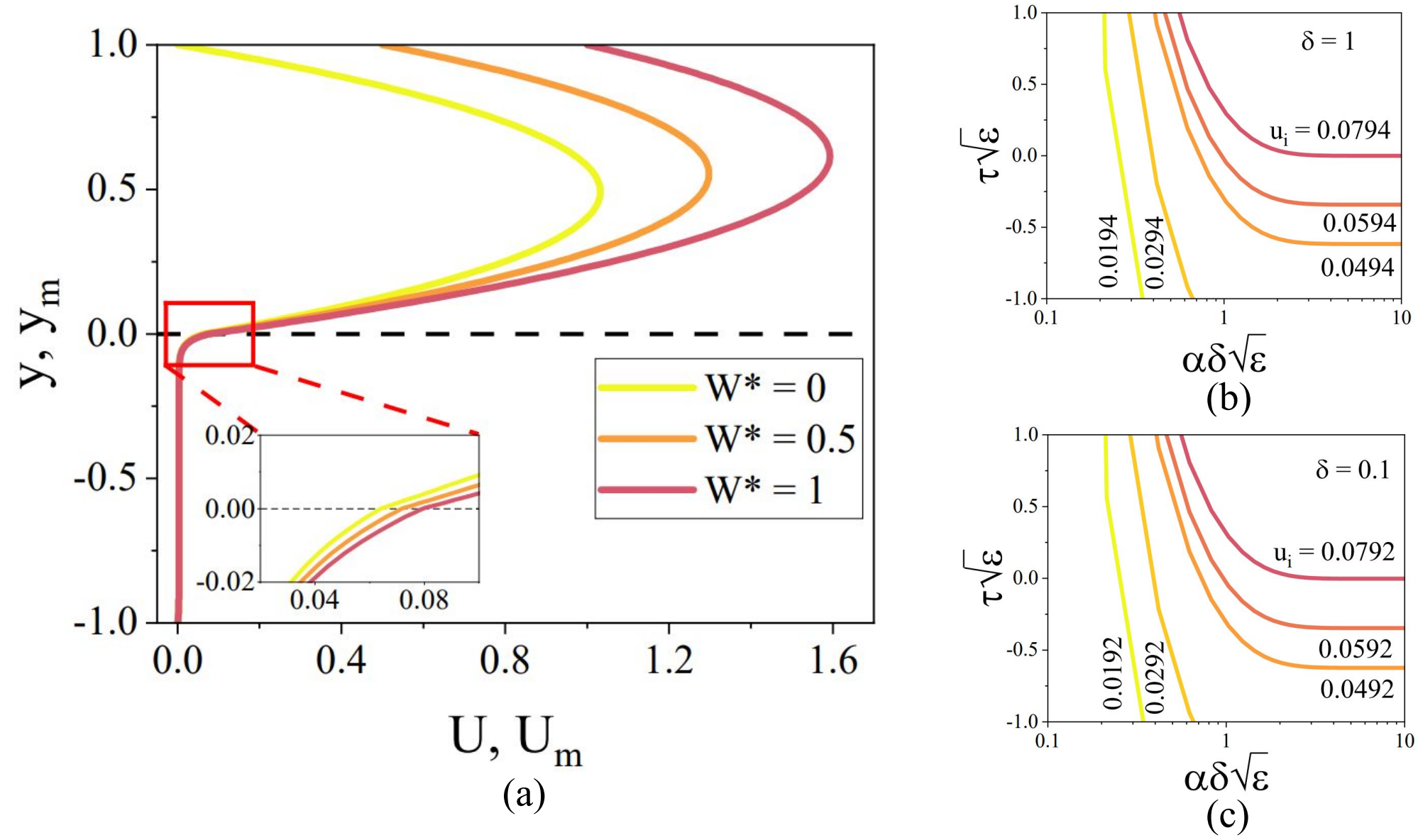

Figure 3: Base state results. (a) The base state velocity profiles using equations 2.12 and 2.13 for different values of Couette component $W^*$. Other parameters are $\epsilon$=0.6, $\tau$=0, $\delta$=1, and $\alpha$=50. (b,c) Contours of normalized slip velocity $u_i$ in the parameter space $(\alpha\delta\sqrt{\epsilon},\tau\sqrt{\epsilon})$ for two different values of $\delta$ (b) $W^*$=1, $\delta$=1; (c) $W^*$=1, $\delta$=0.1.

### 2.3. *Linear stability analysis*

To analyze the linear stability analysis of particle-laden Couette-Poiseuille flow over a porous layer, we introduce infinitesimal perturbations (denoted by the symbol $\tilde{}$), to the base flow: $\boldsymbol{u} = \boldsymbol{U} + \tilde{\boldsymbol{u}}$, $\boldsymbol{u}_p = \boldsymbol{U} + \tilde{\boldsymbol{u}}_p$, $\boldsymbol{u}_m = \boldsymbol{U}_m + \tilde{\boldsymbol{u}}_m$, $p = P + \tilde{p}$, $p_m = P_m + \tilde{p}_m$, $n = N + \tilde{n}$. The dimensionless continuity and momentum equations that govern the perturbed fluid flow are expressed as

$$\frac{\partial \tilde{u}}{\partial x} + \frac{\partial \tilde{v}}{\partial y} + \frac{\partial \tilde{w}}{\partial z} = 0, \tag{2.15}$$

$$\frac{\partial \tilde{u}}{\partial t} + U\frac{\partial \tilde{u}}{\partial x} + \frac{dU}{dy}\tilde{v} = -\frac{\partial \tilde{p}}{\partial x} + \frac{1}{Re}\left(\frac{\partial^2 \tilde{u}}{\partial x^2} + \frac{\partial^2 \tilde{u}}{\partial y^2} + \frac{\partial^2 \tilde{u}}{\partial z^2}\right) - \frac{f}{SRe}(\tilde{u} - \tilde{u}_p), \tag{2.16}$$

$$\frac{\partial \tilde{v}}{\partial t} + U\frac{\partial \tilde{v}}{\partial x} = -\frac{\partial \tilde{p}}{\partial y} + \frac{1}{Re}\left(\frac{\partial^2 \tilde{v}}{\partial x^2} + \frac{\partial^2 \tilde{v}}{\partial y^2} + \frac{\partial^2 \tilde{v}}{\partial z^2}\right) - \frac{f}{SRe}(\tilde{v} - \tilde{v}_p), \tag{2.17}$$

$$\frac{\partial \tilde{w}}{\partial t} + U\frac{\partial \tilde{w}}{\partial x} = -\frac{\partial \tilde{p}}{\partial z} + \frac{1}{Re}\left(\frac{\partial^2 \tilde{w}}{\partial x^2} + \frac{\partial^2 \tilde{w}}{\partial y^2} + \frac{\partial^2 \tilde{w}}{\partial z^2}\right) - \frac{f}{SRe}(\tilde{w} - \tilde{w}_p), \tag{2.18}$$

For the particle phase, the perturbed equations are derived by linearizing the governing equations for particle dynamics around the base state. As in fluid flow, the equations are nondimensionalized using the same characteristic scales. The resulting dimensionless continuity and momentum equations for the particle phase can be expressed as

$$\frac{\partial \tilde{n}}{\partial t} + U\frac{\partial \tilde{n}}{\partial x} + \frac{\partial \tilde{u}_p}{\partial x} + \frac{\partial \tilde{v}_p}{\partial y} + \frac{\partial \tilde{w}_p}{\partial z} = 0, \tag{2.19}$$

$$\frac{\partial \tilde{u}_p}{\partial t} + U\frac{\partial \tilde{u}_p}{\partial x} + \frac{dU}{dy}\tilde{v}_p = \frac{1}{SRe}(\tilde{u} - \tilde{u}_p), \tag{2.20}$$

$$\frac{\partial \tilde{v}_p}{\partial t} + U\frac{\partial \tilde{v}_p}{\partial x} = \frac{1}{SRe}(\tilde{v} - \tilde{v}_p), \tag{2.21}$$

$$\frac{\partial \tilde{w}_p}{\partial t} + U\frac{\partial \tilde{w}_p}{\partial x} = \frac{1}{SRe}(\tilde{w} - \tilde{w}_p), \tag{2.22}$$

Finally, the linearized equations governing the flow within the porous layer are given by

$$\frac{\partial \tilde{u}_m}{\partial x_m} + \frac{\partial \tilde{v}_m}{\partial y_m} + \frac{\partial \tilde{w}_m}{\partial z_m} = 0, \tag{2.23}$$

$$\frac{\partial \tilde{u}_m}{\partial t} = -\epsilon\frac{\partial \tilde{p}_m}{\partial x_m} + \frac{1}{Re}\left(\frac{\partial^2 \tilde{u}_m}{\partial {x_m}^2} + \frac{\partial^2 \tilde{u}_m}{\partial {y_m}^2} + \frac{\partial^2 \tilde{u}_m}{\partial {z_m}^2}\right) - \frac{\epsilon\alpha^2}{Re}\tilde{u}_m, \tag{2.24}$$

$$\frac{\partial \tilde{v}_m}{\partial t} = -\epsilon\frac{\partial \tilde{p}_m}{\partial y_m} + \frac{1}{Re}\left(\frac{\partial^2 \tilde{v}_m}{\partial {x_m}^2} + \frac{\partial^2 \tilde{v}_m}{\partial {y_m}^2} + \frac{\partial^2 \tilde{v}_m}{\partial {z_m}^2}\right) - \frac{\epsilon\alpha^2}{Re}\tilde{v}_m, \tag{2.25}$$

$$\frac{\partial \tilde{w}_m}{\partial t} = -\epsilon\frac{\partial \tilde{p}_m}{\partial z_m} + \frac{1}{Re}\left(\frac{\partial^2 \tilde{w}_m}{\partial {x_m}^2} + \frac{\partial^2 \tilde{w}_m}{\partial {y_m}^2} + \frac{\partial^2 \tilde{w}_m}{\partial {z_m}^2}\right) - \frac{\epsilon\alpha^2}{Re}\tilde{w}_m. \tag{2.26}$$

The corresponding boundary conditions for the perturbed system can be reformulated as

$$\tilde{u} = \tilde{v} = \tilde{w} = 0 \qquad \text{at} \qquad y = 1, \tag{2.27}$$

$$\tilde{u} = \tilde{u}_m, \quad \tilde{v} = \tilde{v}_m, \quad \tilde{w} = \tilde{w}_m \qquad \text{at} \qquad y = 0, \tag{2.28}$$

$$\tilde{p}_m - \frac{2}{\epsilon Re}\frac{\partial \tilde{v}_m}{\partial y_m} = \tilde{p} - \frac{2}{Re}\frac{\partial \tilde{v}}{\partial y} \qquad \text{at} \qquad y = 0, \tag{2.29}$$

$$\frac{1}{\epsilon}\frac{\partial \tilde{u}_m}{\partial y_m} - \frac{\partial \tilde{u}_m}{\partial y_m} = \tau\alpha\tilde{u}_m \qquad \text{at} \qquad y = 0, \tag{2.30}$$

$$\frac{1}{\epsilon}\frac{\partial \tilde{w}_m}{\partial y_m} - \frac{\partial \tilde{w}_m}{\partial y_m} = \tau\alpha\tilde{w}_m \qquad \text{at} \qquad y = 0, \tag{2.31}$$

$$\tilde{u}_m = \tilde{v}_m = \tilde{w}_m = 0 \qquad \text{at} \qquad y = -\delta, \tag{2.32}$$

In the above equations 2.15 - 2.32, several new dimensionless parameters are introduced. The Reynolds number, denoted as $Re$, is defined as $\rho U_{\max} L/\mu$ and is based solely on the pressure-driven velocity scale, while the relative strength of wall motion is characterized independently by $W^*$. The parameter $f$ is expressed as $4\pi a^3 \rho_p N_0/3\rho$, where $a$ is the particle radius, $\rho_p$ is the particle density, $N_0$ is the number of particles per unit volume, and $\rho$ denotes the fluid density. Lastly, $S$ is defined as $2\rho_p a^2/9\rho L^2$. In dusty gas model, particle loading is characterized by the mass fraction $f$ and relaxation time $S$. The corresponding volume fraction is $\phi_{bulk} = f/(1 + \frac{\rho_p}{\rho_f}) \approx f\frac{\rho_f}{\rho_p}$ when $\rho_p \gg \rho$ (heavy particles, as assumed here). For typical density ratios $\rho_p/\rho \approx 1000$ or larger, $\phi_{bulk}$ remains very small even for moderate $f$, ensuring the dilute volume regime. This justifies the pressureless continuum treatment and Stokes drag dominance without the need for granular closures or explicit particle diameter specification.

By assuming a three-dimensional perturbation which are elementary Fourier modes of the form $\tilde{\xi}(x, y, z, t) = \hat{\xi}(y)e^{i(k_x x + k_z z - ct)}$, where $\hat{\xi}(y) = (\hat{u}, \hat{v}, \hat{p}, \hat{n}, \hat{u_m}, \hat{v_m}, \hat{p_m})$, the eigenvalue $c = c_r + ic_i$ is the complex wave speed of perturbations and $k_x$ and $k_z$ are the dimensionless wavenumber. The growth of the perturbation depends on the value of $c_i$, with $c_i > 0$ leading to exponential growth and instability. Using Squire's theorem (see Appendix C), it is shown that two-dimensional (2D) infinitesimal disturbances attain instability at Reynolds numbers that are lower than fully 3D disturbances. Accordingly, the most unstable modes are 2D, and it is sufficient to confine the modal linear stability analysis of particle-laden Couette-Poiseuille flow over a porous substrate to 2D perturbations ( $k_x = k$, $k_z = 0$).

### 2.4. *Numerical method*

The governing equations, together with the associated boundary and interfacial conditions, are transformed into a generalized eigenvalue problem. Spatial discretization is performed using the Chebyshev spectral collocation method, in which the perturbation amplitude functions are represented by truncated expansions of Chebyshev polynomials. Note that each subdomain is discretized independently, the fluid-porous interface is treated as a boundary for both regions, resulting in clustering of collocation points from both sides and ensuring adequate resolution of the interfacial gradients. Since Chebyshev polynomials are defined on the interval $[-1, 1]$, coordinate transformations are applied to map the physical subdomains onto this computational domain following the two- domain approach of Samanta (2020). The fluid region y $\in$ [0,1] and porous region y $\in$ [-$\delta$,0] are discretized using independent Chebyshev collocation grids. Each physical subdomain is mapped separately onto the standard computational interval $[-1, 1]$ using linear transformations.

The resulting discretized system yields a generalized matrix eigenvalue problem, which is solved using the QZ algorithm (Dongarra *et al.* 1996; Samanta 2020; Mirbod *et al.* 2024). Spectral convergence was assessed by increasing the number of Chebyshev collocation points until the relative change in the leading eigenvalues between successive refinements was less than $10^{-4}$, consistent with standard convergence criteria reported in the literature (Tilton & Cortelezzi 2008; Samanta 2017). For cases without the Couette component, 300 collocation points were sufficient to ensure convergence of the dominant eigenvalues and neutral curves. In the presence of the Couette component, a higher resolution (700 collocation points) was required (Samanta 2020). The imposed upper wall motion enlarges the momentum diffusion region near the fluid-porous interface, producing a sharper transition zone that must be accurately resolved. The numerical formulation is validated by comparison with two established limiting cases. First, the model reduces to particle-laden Poiseuille flow in a rigid channel of length $2L$ without a porous layer, for which stability results are available in Klinkenberg *et al.* (2011). Second, in the absence of particles, the formulation recovers single-phase flow over a porous substrate, and the results are compared with those of Samanta (2020). Neutral stability curves obtained in the present study for these limiting scenarios are compared with the published results, as shown in figure 4. The close agreement observed in both cases confirms the correctness of the numerical implementation and demonstrates the model's ability to accurately capture the relevant hydrodynamic stability mechanisms.

## 3. Onset of the modified wall-mode instability

This study presents a systematic linear stability analysis of particle-laden Couette-Poiseuille flow over porous walls, emphasizing how wall permeability, particle relaxation time, and particle mass fraction collectively influence flow stability. The investigation begins with a Poiseuille flow analysis (i.e., $W^* = 0$) to establish a baseline before introducing shear asymmetry. Three representative particle relaxation times ($S = 10^{-7}$, $5 \times 10^{-5}$, and $2.5 \times 10^{-4}$) and particle mass fractions ($f = 0$ to 0.15) are selected based on the framework developed in Klinkenberg *et al.* (2011), enabling controlled parametric variation and comparison with impermeable channel across a wide spectrum of particle inertia effects. We note that the dependence of flow stability on particle relaxation time is generally non-monotonic in particle-laden shear flows, with intermediate values often producing the strongest stabilization. The focus of the present analysis is therefore not on establishing this behavior, but on examining how it is modified by the presence of a porous boundary. The formulation involves several dimensionless parameters: depth ratio $\delta$, porosity $\epsilon$, permeability parameter $\alpha$, momentum transfer coefficient $\tau$, particle relaxation time $S$, mass fraction $f$, Couette ratio $W^*$, and Reynolds number $Re$. To isolate the interplay between particle-fluid coupling and porous-wall effects while maintaining comparability to prior benchmarks (Chang *et al.* 2017; Samanta 2020; Mirbod *et al.* 2024), we fix $\delta = 1$, $\epsilon = 0.6$, $\tau = 0$. We systematically vary $\alpha$, $W^*$, $f$, $S$, and $Re$ (stability curves). This focus allows unambiguous attribution of trends to the particle-porous interaction, revealing permeability-modulated stabilization and Couette effects on critical conditions.

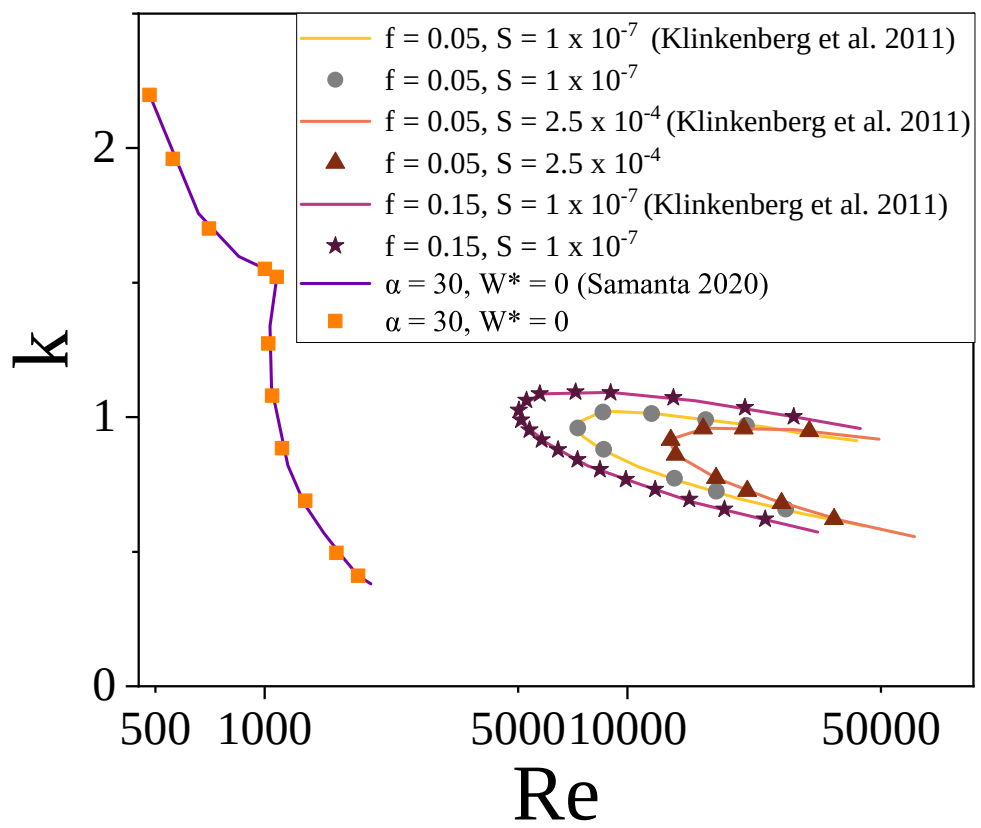

Figure 4: Validation of the numerical model showing good agreement between the current neutral stability curve and benchmark results for particle-laden Poiseuille flow in a rigid channel (Klinkenberg *et al.* 2011) and single-phase flow over a porous substrate (Samanta 2020).

### 3.1. *Particle-laden eigenspectra over porous layers with and without Couette shear*

To clarify the role of wall permeability in shaping both the base flow and its spectral stability, we first determine neutral stability curves for single-phase (particle-free) flows. From these, the critical Reynolds numbers ($Re_{cr}$) and wavenumbers ($k_{cr}$) corresponding to each permeability parameter $\alpha$ are extracted. These benchmarks provide a consistent reference and ensure that the eigenspectra are computed just below the single-phase instability threshold. This choice maximizes sensitivity to perturbations, allowing us to isolate and identify new unstable branches that emerge when particles are introduced. We focus on three permeability values, $\alpha = 50, 100$ and 400, spanning the regime from moderately to weakly permeable walls. Since the permeability parameter $\alpha$ is inversely proportional to $\sqrt{\kappa}$, $\alpha = 50$ represents the most permeable configuration considered in this study, while $\alpha = 400$ represents the least permeable (quasi-impermeable) wall.

Figure 5 shows eigenspectra for flows over porous walls with $\alpha = 100$ and $\alpha = 400$, comparing three particle mass fractions ($f = 0$, 0.05, and 0.15) across three relaxation times. For $\alpha = 100$, simulations are performed at $Re = 1550$ and $k = 2.7$, while for $\alpha = 400$, they are carried out at $Re = 6500$ and $k = 2.18$. These parameter sets lie just below the neutral stability threshold for particle-free flow, ensuring that the base state remains stable in the absence of suspended particles and thereby isolating particle-induced modifications to the eigenvalue spectrum. Figure 5a presents the eigenspectra of a fluid flow laden with particles with a particle relaxation time of $S = 1 \times 10^{-7}$, considering varying mass fractions $f$, in the presence of a porous layer characterized by $\alpha = 100$ within the channel. The eigenspectrum exhibits a distinctive Y-shaped structure, characterized by the distribution of eigenvalues across three clearly identifiable branches (Schmid *et al.* 2002). The presence of permeability introduces additional eigenvalues into the eigenspectrum, referred to as porous modes. These new modes arise due to significant variations in the eigenfunctions near the porous layer, as discussed in previous studies (Tilton & Cortelezzi 2008). We emphasize that these additional porous modes remain stable; however, their presence modifies the classical fluid modes through interfacial coupling, leading to shifts in growth rates and, in some cases, altering which mode governs the onset of instability.

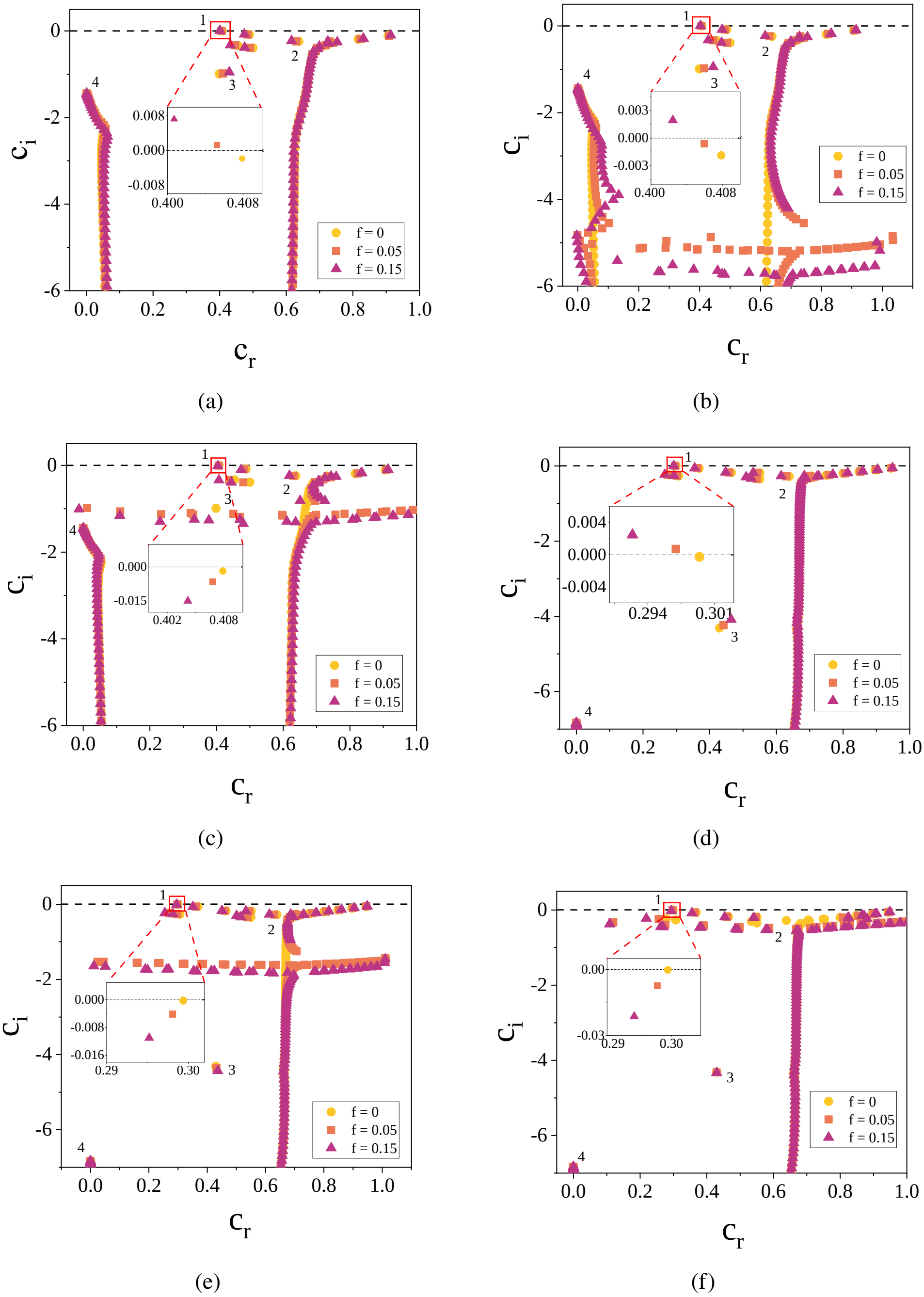

Figure 5: Eigenspectra variation for different particle parameters (relaxation time S and mass fraction f) and permeability parameter $\alpha$ for flow without Couette component ($W^* = 0$). (a) $\alpha$=100, $S = 1 \times 10^{-7}$; (b) $\alpha$=100, $S = 5 \times 10^{-5}$; (c) $\alpha$=100, $S = 2.5 \times 10^{-4}$; (d) $\alpha$=400, $S = 1 \times 10^{-7}$; (e) $\alpha$=400, $S = 5 \times 10^{-5}$; (f) $\alpha$=400, $S = 2.5 \times 10^{-4}$. Other parameters: $Re$ = 1550, $k$ = 2.7 for $\alpha$=100, $Re$ = 6500, $k$ = 2.18 for $\alpha$=400, $\tau$=0, $\epsilon$=0.6, $\delta$=1.

To analyze the impact of particles, we examine the emergence and trajectories of four distinct modes, labeled 1-4 in figure 5a. Modes 1 and 2 belong to the wall mode branch, with Mode 1 (porous-modified Tollmien-Schlichting mode (TSM)) exhibiting the highest growth rate among all modes, while Modes 3 and 4 emerge due to the influence of the porous layer. Since the chosen values of $Re$ and $k$ lie below the critical thresholds for plane Poiseuille flow ($f$=0) with $\alpha$=100, the maximum growth rate ($c_i$) remains negative but close to zero, while all other eigenvalues also exhibit negative $c_i$, indicating a stable system.

Interestingly, the general structure of the eigenspectra remains largely unchanged with the addition of particles. However, the porous-modified TS mode, which was initially stable in the absence of particles, becomes unstable when particles with a relaxation time of $S = 1 \times 10^{-7}$ are introduced. Specifically, Mode 1 transitions to an unstable state at $f = 0.05$, with further destabilization as $f$ increases to 0.15. Similarly, the growth rates of Modes 2 and 3 increase with rising $f$, yet they remain below $c_i = 0$, indicating their stability. In contrast, Mode 4 remains largely unaffected by particle addition, suggesting that while particle-fluid interactions influence the overall stability of the flow, their impact on the porous mode is minimal. This observation implies that particles primarily affect the stability of wall modes, while the characteristics of the porous modes remain largely unaltered.

The destabilization of the flow due to the addition of particles with a relaxation time of $S = 1 \times 10^{-7}$ has been observed even in the absence of a porous layer (refer to figure 13b in A.1). This destabilizing effect is reflected in a reduction of the critical Reynolds number, indicating an earlier transition to instability compared to single-phase flow. Similar behavior has been reported in the neutral stability curve analysis by Klinkenberg *et al.* (2011), whereas in this work we specifically characterize it by analyzing the $c_i - c_r$ curve, highlighting the consistent influence of small inertia particles on flow destabilization across different flow configurations. For such small relaxation times ($S = 1 \times 10^{-7}$), the particles are extremely small and behave almost as passive tracers. Consequently, the primary effect of these particles is an increase in the system's overall density, which effectively lowers the critical Reynolds number by a factor, thereby promoting instability (Saffman 1962). In the present configuration, however, this mechanism interacts with the porous boundary, leading to a modified stability response compared to impermeable-wall flows.

Similarly, as the particle mass fraction $f$ increases, the growth rate of porous-modified TS mode rises for particles with a relaxation time of $S = 5 \times 10^{-5}$, leading to a pronounced destabilization effect as shown in figure 5b. The variation in the growth rates of Modes 2 to 4 follows a similar trend to the case with $S = 1 \times 10^{-7}$ (see figure 5a). However, due to the higher relaxation time, particles with $S = 5 \times 10^{-5}$ exhibit more pronounced lag effects in their response to fluid velocity fluctuations. This lag results in the emergence of distinctive Y-shaped structures in the eigenspectrum, characterized by low $c_i$ values and the presence of additional damped modes. These new modes arise due to increased particle-fluid interactions, a consequence of the longer relaxation time, and are distinct from the porous modes associated with the underlying permeable layer. A key observation here is that while these newly formed Y-shaped structures become more pronounced with increasing $f$, their growth rates remain lower than those of the primary unstable modes. This indicates that although additional modes are generated, they do not significantly contribute to the overall flow instability. Instead, the destabilizing effect is primarily driven by the porous-modified TS mode, whose growth rate increases markedly with higher mass fractions, as observed in figure 5b.

In the absence of a porous layer, the addition of particles with a relaxation time of $S = 5{\times}10^{-5}$ exhibits a contrasting stabilizing effect. As shown in figure 13c in A.1, increasing the mass fraction $f$ results in a reduction in the leading eigenvalues. This stabilization is characterized by a shift in the neutral stability curve toward higher critical Reynolds numbers, indicating that a higher flow velocity is required for the onset of instability Klinkenberg *et al.* (2011). However, the introduction of a porous layer alters this stabilizing behavior. With the porous substrate present, the system transitions to a destabilizing trend as $f$ increases. This transition is driven by the interaction between the porous layer and the particle-laden fluid, in which the porous layer acts as a catalyst for the amplification of disturbances. This dual effect, stabilization in the absence of a porous layer and destabilization in its presence, highlights the complex interplay between particle dynamics, flow structure, and interactions of porous media.

For particles with a higher relaxation time of $S = 2.5 \times 10^{-4}$, the increase in mass fraction $f$ produces a stabilizing effect on the flow. As shown in figure 5c, the growth rate of the porous-modified TS mode consistently decreases with increasing $f$, indicating a suppression of instability. Notably, despite the emergence of a distinct Y-shaped structure in the eigenspectrum for $S = 2.5 \times 10^{-4}$, the growth rates of these new modes remain below zero, confirming that they do not actively contribute to instability. This trend contrasts with the behavior observed for smaller relaxation times, where particle-fluid interactions tended to amplify the porous-modified TS mode, leading to destabilization. The stabilization effect associated with $S = 2.5 \times 10^{-4}$ can be attributed to the substantial inertial lag of these heavier particles, which dissipates flow disturbances more effectively than particles with lower relaxation times. As a result, the influence of the porous layer is effectively dampened, leading to a more stable flow configuration. This stabilizing effect is further reinforced in the absence of a porous layer, as evidenced by the monotonic decrease in the growth rate of the TS mode with increasing $f$, as illustrated in figure 13d in A.1. Here, the neutral stability curve shifts toward higher critical Reynolds numbers, mirroring the behavior reported by Klinkenberg *et al.* (2011) and indicating that larger flow velocities are required to induce instability.

Increasing the permeability parameter $\alpha$ to 400 results in a pronounced reduction in permeability, effectively transforming the porous layer into a quasi-rigid boundary. This transition is apparent in figure 5d, where the porous mode diminishes in prominence, signifying the decreased influence of the porous layer on flow dynamics. For small particles with a relaxation time of $S = 1 \times 10^{-7}$, increasing $f$ from 0 to 0.15 leads to a pronounced destabilization, characterized by an increase in the growth rate of the porous-modified TS mode. This trend is further compounded by the simultaneous decline in Mode 3's growth rate as $\alpha$ increases, highlighting the stabilizing effect of reduced permeability. However, the response of particles with higher relaxation times, specifically $S = 5{\times}10^{-5}$ and $S = 2.5{\times}10^{-4}$, diverges significantly. As illustrated in figures 5e and 5f, the growth rate of the porous-modified TS mode decreases with increasing $f$, indicating a stabilizing effect. Notably, for $S = 2.5\times10^{-4}$, the growth rate of the newly formed Y-shaped structures increases but remains below zero, affirming that these structures do not induce instability. The observed stabilization with $\alpha = 400$ suggests that the porous layer's influence diminishes as its permeability is reduced. Furthermore, as $f$ increases, the system exhibits a marked suppression of the most unstable modes, underscoring the critical role of particle relaxation time in dictating flow stability over porous substrates. This interplay between particle dynamics, porous media, and flow stability is further examined for varying $\alpha = 50$ in A.2, where increasing the particle mass fraction $f$ leads to a consistent destabilization across all values of the relaxation parameter $S$.

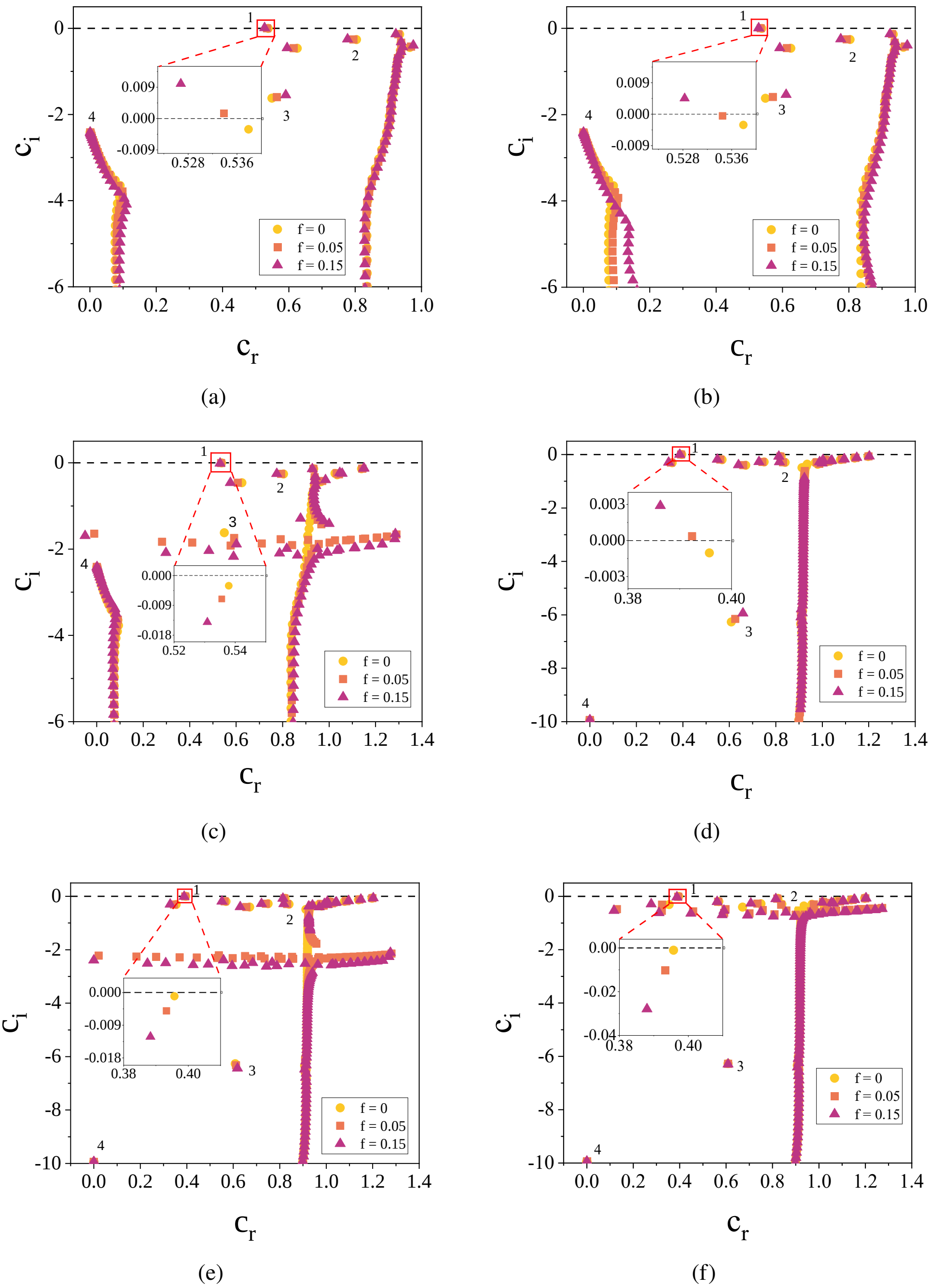

Figure 6: Eigenspectra variation for different particle parameters (relaxation time S and mass fraction f) and permeability parameter $\alpha$ for flow with Couette component ($W^* = 0.5$). (a) $\alpha$=100, $S = 1 \times 10^{-7}$; (b) $\alpha$=100, $S = 5 \times 10^{-5}$; (c) $\alpha$=100, $S = 2.5 \times 10^{-4}$; (d) $\alpha$=400, $S = 1 \times 10^{-7}$; (e) $\alpha$=400, $S = 5 \times 10^{-5}$; (f) $\alpha$=400, $S = 2.5 \times 10^{-4}$. Other parameters: $Re = 1250$, $k = 2$ for $\alpha$=100, $Re = 7000$, $k = 1.38$ for $\alpha$=400, $\tau$=0, $\epsilon$=0.6, $\delta$=1.

For particles with very small relaxation time (e.g., $S = 1 \times 10^{-7}$), the particle velocity rapidly adjusts to the fluid velocity, and the relative slip between phases is negligible. Their primary effect is to modify the effective inertia of the mixture, thereby shifting the critical Reynolds number. In contrast, for particles with larger relaxation time ($S = 5 \times 10^{-5}$, $2.5 \times 10^{-4}$), the lag induces significant slip in perturbations, leading to energy extraction from the disturbance field and the emergence of additional coupled modes, which are stable and do not contribute to instability. The stable modes arise precisely from the inclusion of relative slip during perturbations, which provides dissipation without introducing new unstable mechanisms in this uniform-distribution, parallel shear setup. Destabilization typically requires particle concentration gradients or inhomogeneities to create feedback loops (e.g., Boronin (2012); Boronin & Osiptsov (2014); Kumar & Govindarajan (2024)), absent here.

We next assess the influence of the Couette component, characterized by non-zero $W^*$, on the linear stability of particle-laden flows over porous layers. Figure 6 presents eigenspectra for flows above porous substrates with $\alpha = 100$ and $\alpha = 400$, in the presence of a Couette component $W^* = 0.5$. For these cases, the spectra are evaluated at subcritical parameters: $Re = 1250$, $k = 2$ for $\alpha = 100$, and $Re = 7000$, $k = 1.38$ for $\alpha = 400$, ensuring that the base flows remain stable in the absence of particles. The eigenspectra retain the characteristic porous modes and the distinct Y-shaped structure observed previously, composed of wall, center, and damped branches. This persistence indicates that the fundamental classification of modes is preserved despite the introduction of shear. The Y-shaped structure again reflects spatial localization: wall modes are concentrated near the boundaries, center modes dominate in the channel core, and damped modes remain weakly amplified. Since wall modes are most sensitive to near-boundary dynamics, the addition of Couette shear has its strongest effect on these disturbances. As shown in figure 6, instability onset continues to originate from Mode 1 (the porous-modified TS mode). The imposed wall shear amplifies this branch more strongly than either center or damped modes, underscoring the enhanced receptivity of near-wall instabilities to Couette-driven shear.

Figures 6a to 6c illustrate the influence of Couette flow on the eigenspectrum for particles with relaxation times of $S = 1 \times 10^{-7}$, $5 \times 10^{-5}$, and $2.5 \times 10^{-4}$, at a fixed permeability parameter of $\alpha$=100. The observed trends in the evolution of Modes 1 through 4 align with those observed in the absence of the Couette component (as shown in Figure 5), suggesting that while the Couette component modifies the growth rates, the fundamental mode structure remains intact. Similarly, figures 6d to 6f present the eigenspectra for $\alpha$=400, under the same set of particle relaxation times. Here too, the variation in mode structures is consistent with that observed in the absence of Couette flow, further affirming that while shear modifies the growth rates of certain modes, the overall structure of the eigenspectrum remains largely unaffected.

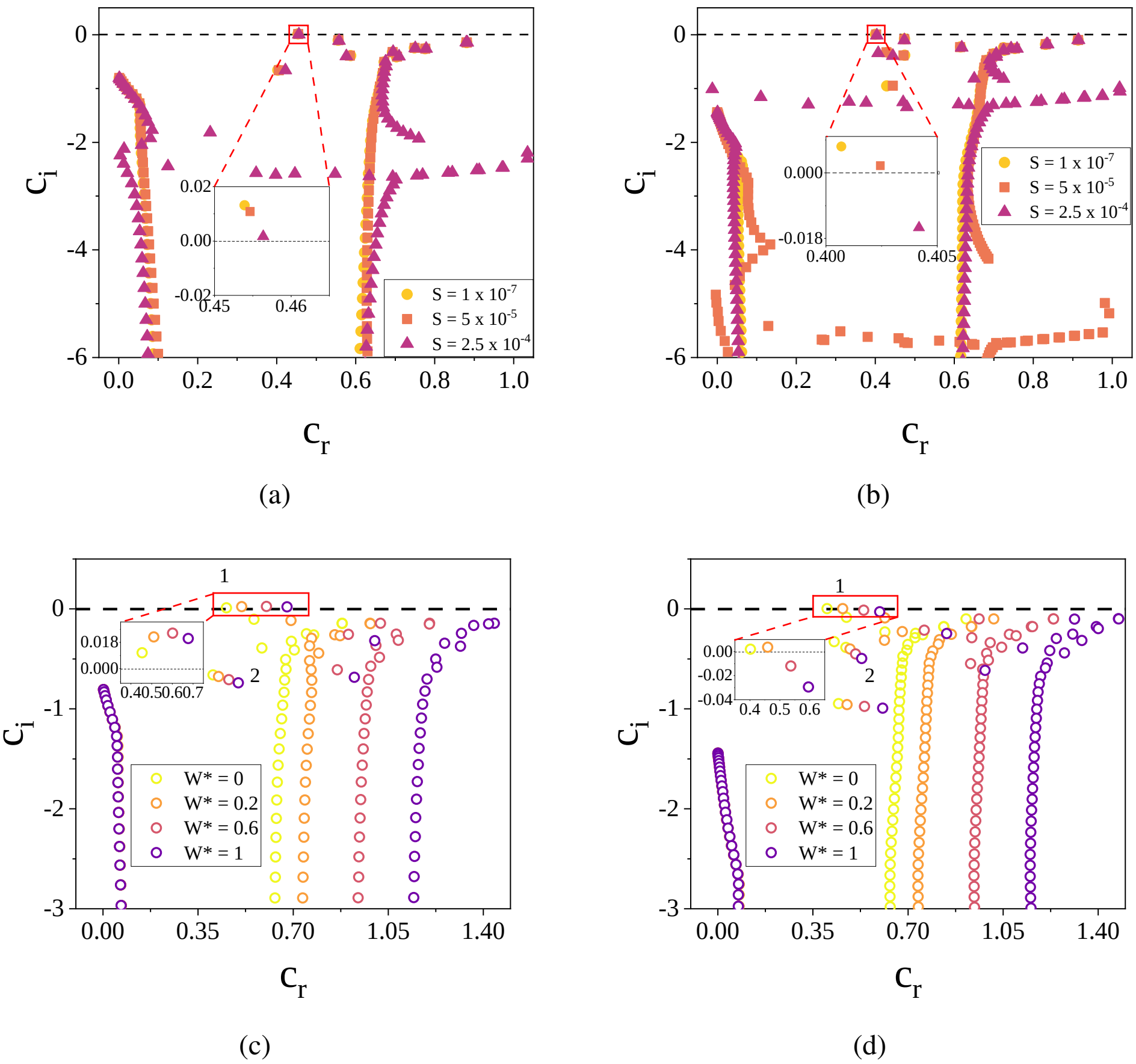

Figure 7: Eigenspectra variation for different particle parameters (relaxation time S) and permeability parameter $\alpha$. (a) $\alpha$=50, $f$=0.15, $W^*$=0, $Re$=680, $k$=2.8; (b) $\alpha$=100, $f$=0.15, $W^*$=0, $Re$=1550, $k$=2.7; (c) $\alpha$=50, $f$=0.15, $S = 5 \times 10^{-5}$, $Re$=680, $k$=2.8; (d) $\alpha$=100, $f$=0.15, $S = 5 \times 10^{-5}$, $Re$=1550, $k$=2.7. Other parameters: $\tau$=0, $\epsilon$=0.6, $\delta$=1.

### 3.2. *Effect of varying $W^*$, S, and f*

To probe the interplay between particle relaxation, shear asymmetry, and wall permeability, we first examine the impact of relaxation time $S$ at fixed mass fraction ($f = 0.15$) for moderately permeable walls ($\alpha = 50$ and 100). Figures 7a and 7b show that at $\alpha = 50$, increasing $S$ progressively stabilizes the dominant unstable branch (the porous-modified TS mode) by reducing its growth rate $c_i$, although the instability persists in all $S$. In the limit of very large $S$ ($S \to \infty$), particles become increasingly inertial and unable to respond to fluid perturbations. The drag term proportional to $(u_p - u)/S$ therefore vanishes, the particle phase effectively decouples from the disturbance dynamics. In this limit, the system approaches the corresponding single-phase Couette-Poiseuille flow over a porous medium, which is stable at the considered $Re$ and $k$. This behavior contrasts with impermeable channels, where particles with relaxation times $S = 5 \times 10^{-5}$ and $2.5 \times 10^{-4}$ are associated with a stabilizing effect over the explored Reynolds number range, highlighting the destabilizing influence of wall permeability (Klinkenberg *et al.* 2011). In $\alpha = 100$, the same stabilizing trend is observed, but weaker permeability promotes greater damping; in particular, for $S = 2.5 \times 10^{-4}$, the instability is nearly quenched, suggesting that larger particle inertia can counteract porous-induced destabilization at reduced permeability.

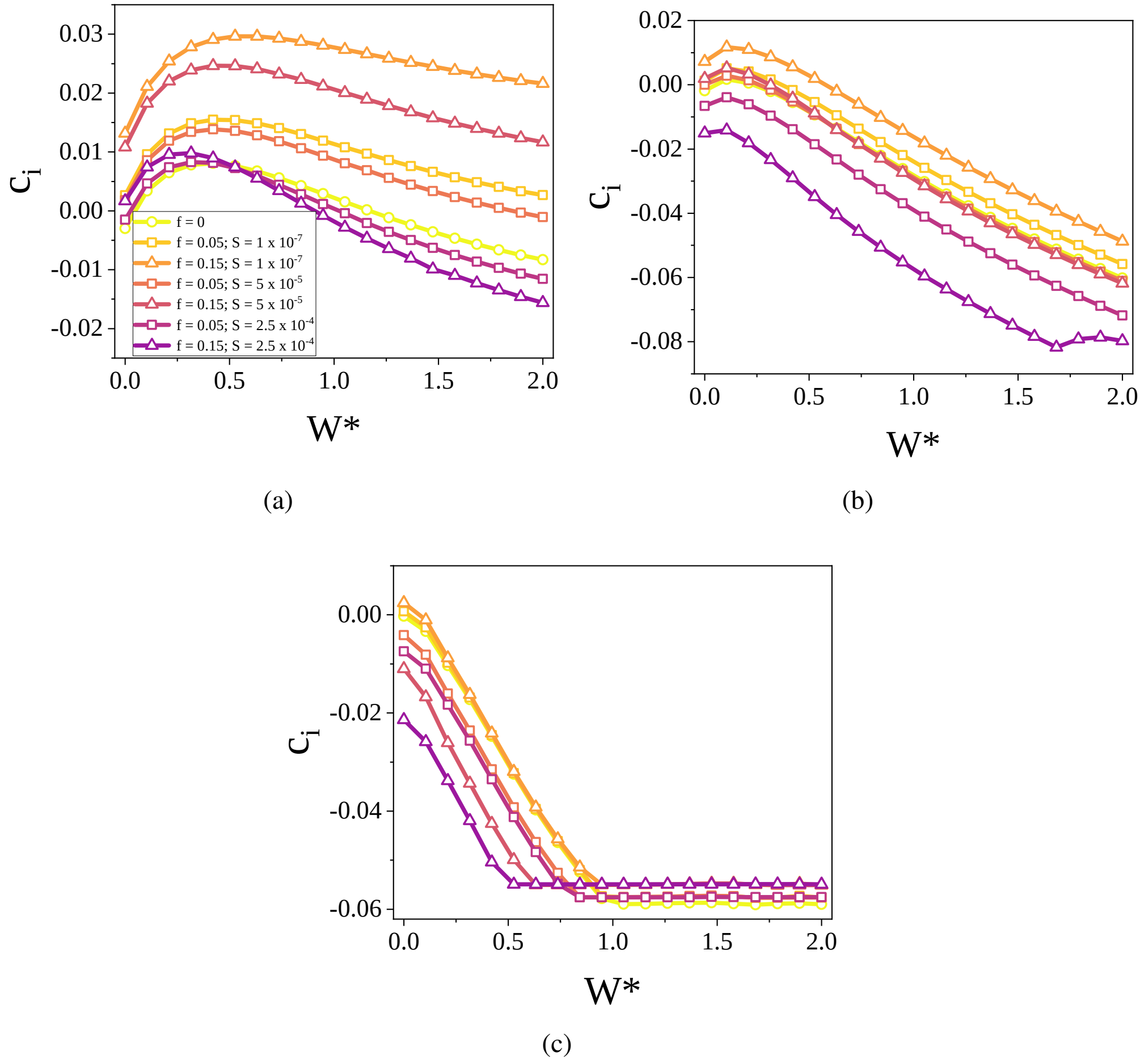

Figure 8: Variation of growth rate with Couette component for different mass fraction ($f$), relaxation time ($S$), and permeability parameter $\alpha$. (a) $\alpha$=50, $k$=2.8, $Re$=680; (b) $\alpha$=100, $k$=2.7, $Re$=1550; (c) $\alpha$=400, $k$=2.18, $Re$=6450. Other parameters: $\tau$=0, $\epsilon$=0.6, $\delta$=1.

Figures 7c and 7d examine the influence of increasing Couette component $W^*$ for fixed $f = 0.15$ and $S = 5 \times 10^{-5}$ at $\alpha = 50$ and $\alpha = 100$, respectively. In both cases, the dominant unstable mode (Mode 1) emerges from the wall region and shows an increasing growth rate with increasing $W^*$ to a peak, after which it slowly decays. Despite this reduction, Mode 1 remains the primary driver of instability. Meanwhile, Mode 2, which is marginally stable at $W^* = 0$, is further stabilized with increasing $W^*$, and eventually becomes more strongly damped. The eigenspectrum collectively shifts rightward in the complex plane as $W^*$ increases, indicating a redistribution of mode dynamics influenced by asymmetric shear. In particular, the porous modes appear to be largely insensitive to the Couette component, with only slight damping observed at higher $\alpha$.

Figure 8 presents the variation of perturbation growth rate ($c_i$) with Couette shear ($W^*$) at a fixed $Re$ and $k$, for different particle parameters and three permeability regimes: low ($\alpha$ = 50), moderate ($\alpha$ = 100), and high ($\alpha$ = 400). The figures capture the combined influence of wall-induced shear, particle relaxation time ($S$), and mass fraction ($f$) on linear instability in Couette-Poiseuille flow over a porous substrate. Changes in $c_i$ reflect a local modification of instability at the chosen $Re$ and $k$, and should be distinguished from the global stability behaviour characterized by the critical Reynolds number $Re_{cr}$, discussed later. In the low-permeability parameter ($\alpha = 50$, figure 8a), the growth rate exhibits a non-monotonic dependence on $W^*$, with an initial increase followed by a decrease after reaching a peak at a critical shear velocity $W^*_{\rm cr} \approx 0.4$. This behavior reflects the dual role of Couette shear: it initially enhances the wall-mode instability through near-wall shear amplification, but eventually stabilizes the flow by disrupting coherent perturbation structures. The particle dynamics introduces an additional layer of complexity: for small $S$ values (e.g., $S = 10^{-7}, 5 \times 10^{-5}$), increasing $f$ raises $c_i$, enhancing the instability. However, for a longer relaxation time ($S = 2.5 \times 10^{-4}$), the system exhibits a transition: the flow is destabilized for $W^* < W^*_{\rm cr}$ but stabilizes for $W^* > W^*_{\rm cr}$, with $c_i$ falling below that of the particle-free case. These trends align with the findings of Samanta (2020), who observed initial destabilization in Newtonian Couette-Poiseuille flows on porous substrates for small $W^*$ and high $\alpha$.

In the moderate permeability parameter ($\alpha = 100$, figure 8b), the influence of $W^*$ is predominantly stabilizing beyond a narrow initial regime, where a slight increase in growth rate is observed. This small initial rise reflects the transition between weak and moderate Couette forcing, after which further increases in $W^*$ consistently reduce the growth rate. The initial shear amplification observed at $\alpha = 50$ is suppressed due to the reduced permeability-induced wall interaction. For $S = 10^{-7}$ and $S = 5 \times 10^{-5}$, increasing $f$ still raises $c_i$, but the magnitude of destabilization is significantly reduced. In contrast, for $S = 2.5 \times 10^{-4}$, increasing $f$ consistently reduces $c_i$, indicating complete particle-induced stabilization. At a high permeability parameter ($\alpha = 400$, figure 8c), the flow behavior shifts dramatically. Here, the growth rate decreases monotonically with increasing $W^*$ and plateaus for $W^* \gtrsim 1.2$. This indicates that wall shear fully suppresses the instability and that the porous layer effectively behaves as a solid boundary. This observation is consistent with Chang *et al.* (2017), who demonstrated that increasing wall motion leads to a higher critical Reynolds number in Couette-Poiseuille systems, particularly when the wall behaves rigidly(equivalent to $\alpha \gtrsim 1000$ in our framework).

Once the plateau is reached, the effects of particle relaxation and variations in $f$ are largely muted. These results reinforce the observations by Potter (1966) and Hains (1967), who found that beyond a threshold Couette velocity, further increases in wall motion have little effect on flow stability. The findings have implications for the design and control of suspension transport in applications such as filtration, biomedical flows through porous tissues, and additive manufacturing, where tuning wall permeability and particle properties can be leveraged to suppress or promote instabilities.

### 3.3. *Neutral stability curves and critical parameters*

We proceed by analyzing the effect of particle addition on the neutral stability curve in a channel flow overlying a porous layer in the absence of the Couette component($W^*$=0), as shown in figure 9. The neutral curve, which delineates the boundary between stable and unstable regimes in the $(Re, k)$ plane, is calculated for three representative parameters of permeability ($\alpha$=50, 100, 400) over a range of fractions of particle mass ($f$) and relaxation times ($S$).

For low to moderate permeability parameters ($\alpha$ = 50 and 100), the neutral curves exhibit a distinctive bimodal structure, consistent with earlier findings for Newtonian porous wall flows (Samanta 2017; Hooshyar *et al.* 2022). The left lobe of the curve originates in a porous-associated mode, while the right lobe corresponds to a fluid-dominated mode. Local minima within these lobes identify regions of enhanced instability, with the global minimum consistently located in the fluid mode. This indicates that the dominant instability mechanism is governed primarily by fluid dynamics rather than by direct coupling with the porous substrate. As permeability decreases (i.e., as $\alpha$ increases), the influence of the porous mode is progressively attenuated. The left lobe becomes less distinct and eventually merges with the right lobe, yielding a single unimodal neutral curve at $\alpha = 400$. This transition highlights the diminishing role of interfacial momentum exchange as the porous layer approaches a quasi-rigid boundary condition, in line with the stabilization trends reported by Hooshyar *et al.* (2022).

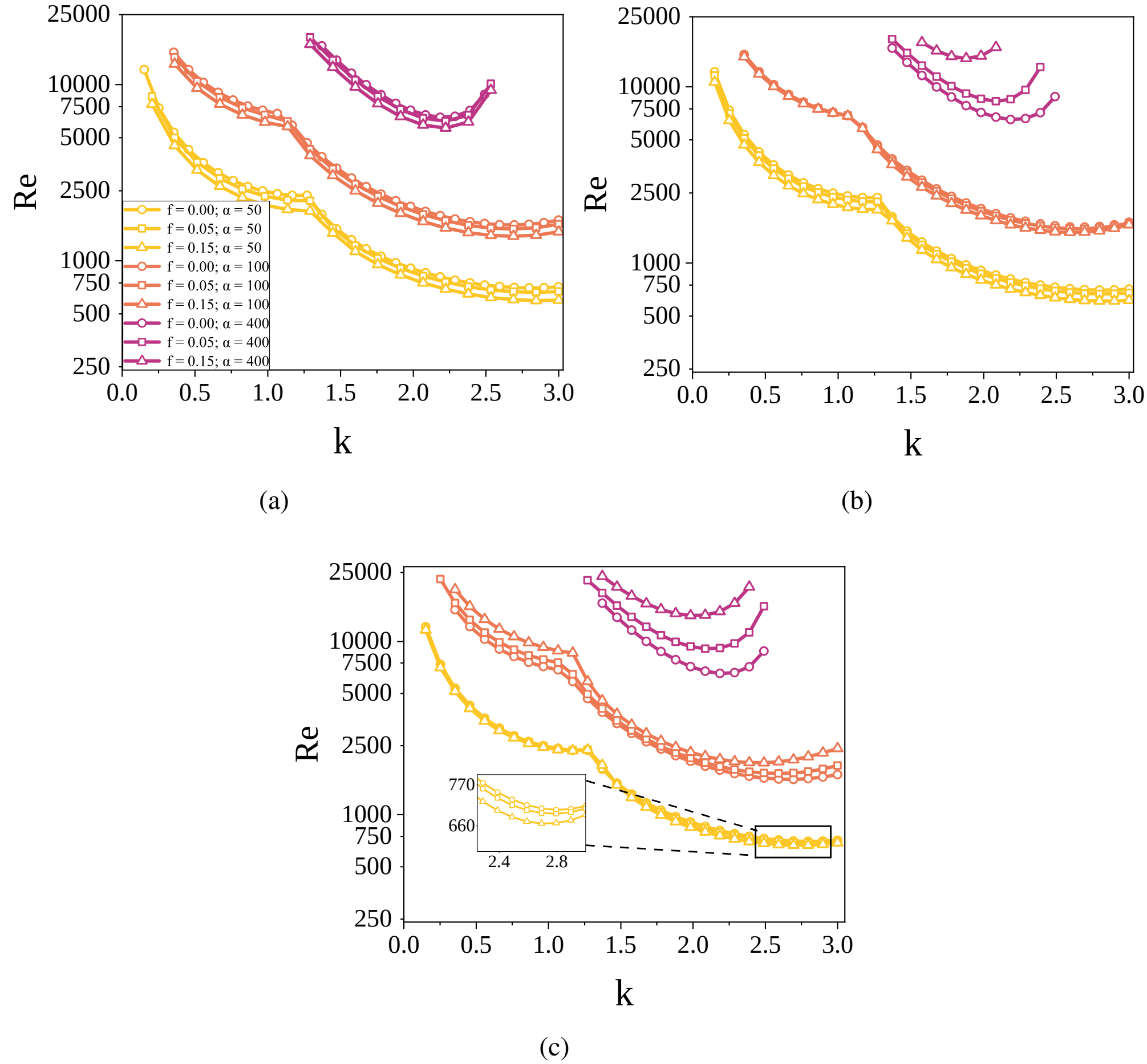

Figure 9: The variation of neutral curve for different values of particle relaxation time $S$, mass fraction $f$, and permeability parameter $\alpha$ without Couette component ($W^* = 0$). (a) $S = 1 \times 10^{-7}$; (b) $S = 5 \times 10^{-5}$; (c) $S = 2.5 \times 10^{-4}$. Other parameters: $\tau$=0, $\epsilon$=0.6, $\delta$=1.

As can be observed in figure 9a, for particles with $S = 1 \times 10^{-7}$, increasing the mass fraction $f$ leads to a consistent reduction in the critical Reynolds number ($Re_{cr}$) for all three values of $\alpha$. This reduction indicates that adding particles enhances the destabilizing effect of the flow, thereby promoting instability. For particles with a longer relaxation time of $S = 5 \times 10^{-5}$, the response of the neutral curve exhibits more nuanced behavior. Increasing the mass fraction $f$ from 0 to 0.15 results in a reduction in $Re_{cr}$ for $\alpha$=50 and $\alpha$=100, consistent with the destabilizing trend observed for $S = 1 \times 10^{-7}$ (as shown in figure 9b). However, for $\alpha$=400, this trend reverses, with $Re_{cr}$ increasing as $f$ increases. This is because at large $\alpha$, the porous layer strongly suppresses flow and the instability structure approaches that of an impermeable wall suspension. In this regime, particles with intermediate relaxation time ($S = 5 \times 10^{-5}$) are sufficiently inertial to lag the fluid disturbances and extract energy through interphase drag, leading to an additional stabilizing influence as the particle loading $f$ increases. For particles with the highest relaxation time of $S = 2.5 \times 10^{-4}$, the influence of particle addition becomes increasingly complex. As depicted in figure 9c, while increasing $f$ leads to a decrease in $Re_{cr}$ for $\alpha$=50, suggesting a destabilizing effect, for $\alpha = 100$ and $\alpha = 400$, $Re_{cr}$ increases with increasing $f$, indicating a pronounced stabilizing effect under these conditions. The contrasting behavior observed for different relaxation times and permeability values underscores the interplay between particle-fluid interactions and the porous substrate. In particular, for intermediate relaxation times ($S = 5 \times 10^{-5}$), increasing the mass fraction $f$ destabilizes the flow for $\alpha = 100$, while stabilizing it for $\alpha = 400$. This dual behavior can be attributed to the relative strengths of particle inertia and porous-layer resistance, as explored in detail through eigenfunction and stream-function analysis presented in Appendix B.

Hence, particle addition does not exert a uniformly stabilizing influence. Instead, the effect on neutral stability curves is highly dependent on both particle relaxation time $S$ and porous permeability parameter $\alpha$. While particle inertia can exert a stabilizing influence in an impermeable wall configurations ($S = 5 \times 10^{-5}$ and $2.5 \times 10^{-4}$), this stabilization weakens or reverses at high permeability (low $\alpha$), where particle addition can even destabilize the system relative to the unladen case. These results highlight a central novel finding: the classical particle-induced stabilization observed in pure suspension flows over impermeable walls can be overturned by the presence of a porous substrate. Destabilization dominates at high permeability and/or low particle inertia, whereas pronounced stabilization emerges at low permeability and higher inertia, reflecting the competition between porous-induced flow penetration (destabilizing) and particle damping via Stokes drag (stabilizing).

The effect of the Couette component on flow stability is further analyzed by examining the neutral curve for a fixed Couette velocity ratio, $W^* = 0.5$. The corresponding neutral curves are depicted in figures 10a and 10b, which illustrate the effect of varying $\alpha$ and $f$ at different $S$. In the presence of the Couette component, the neutral curve consistently adopts a unimodal structure, regardless of the permeability or particle parameters. This unimodal behavior contrasts with the bimodal structure observed in the absence of a Couette component, particularly for lower values of $\alpha$. The primary instability mode is now confined to the fluid mode, with the influence of the porous mode significantly diminished.

A key observation in the presence of Couette flow is the consistent reduction in $Re_{cr}$ for the dominant mode (fluid mode), as compared to the zero Couette component case. This reduction in $Re_{cr}$ signifies a pronounced destabilizing effect, indicating that the velocity of the upper wall exerts a destabilizing influence on the system. This destabilizing trend is observed across all examined values of $f$ and $S$. Specifically, the increased upper wall velocity enhances momentum transfer into the fluid layer, thereby intensifying flow disturbances and promoting instability. When examining the effect of particle addition in the Couette-driven flow, the variation in the neutral curve exhibits a trend analogous to that observed in the zero Couette case, but with a consistently lower $Re_{cr}$. The reduction in $Re_{cr}$ underscores the enhanced destabilizing effect induced by the combination of particle-laden flow and Couette shear. Notably, the destabilization effect becomes more pronounced with increasing particle mass fraction $f$, highlighting the compounded impact of particle-induced inertia and shear-driven momentum transfer.

A deeper understanding of the instability mechanism can be gained by examining the profiles of normalized eigenfunctions and the corresponding contour patterns of the stream function. As shown in figure 10c, the introduction of a Couette component produces pronounced variations in the eigenfunction within the central region of the fluid layer. This is due to the dominance of the fluid-layer mode in the Couette component, which promotes enhanced disturbance amplification and the development of a stronger vortex near the channel center, as shown in figure 10d. This intensified vortex activity is more pronounced than that observed in the absence of Couette flow, as demonstrated in figure 15 (Appendix B). The intensified vortex structures indicate a more vigorous mixing at the fluid-porous interface, driven by the enhanced momentum diffusion caused by the upper wall motion. This effect is consistent with the findings of Samanta (2020), who reported a similar mechanism of shear-induced destabilization in Couette-driven porous flows.

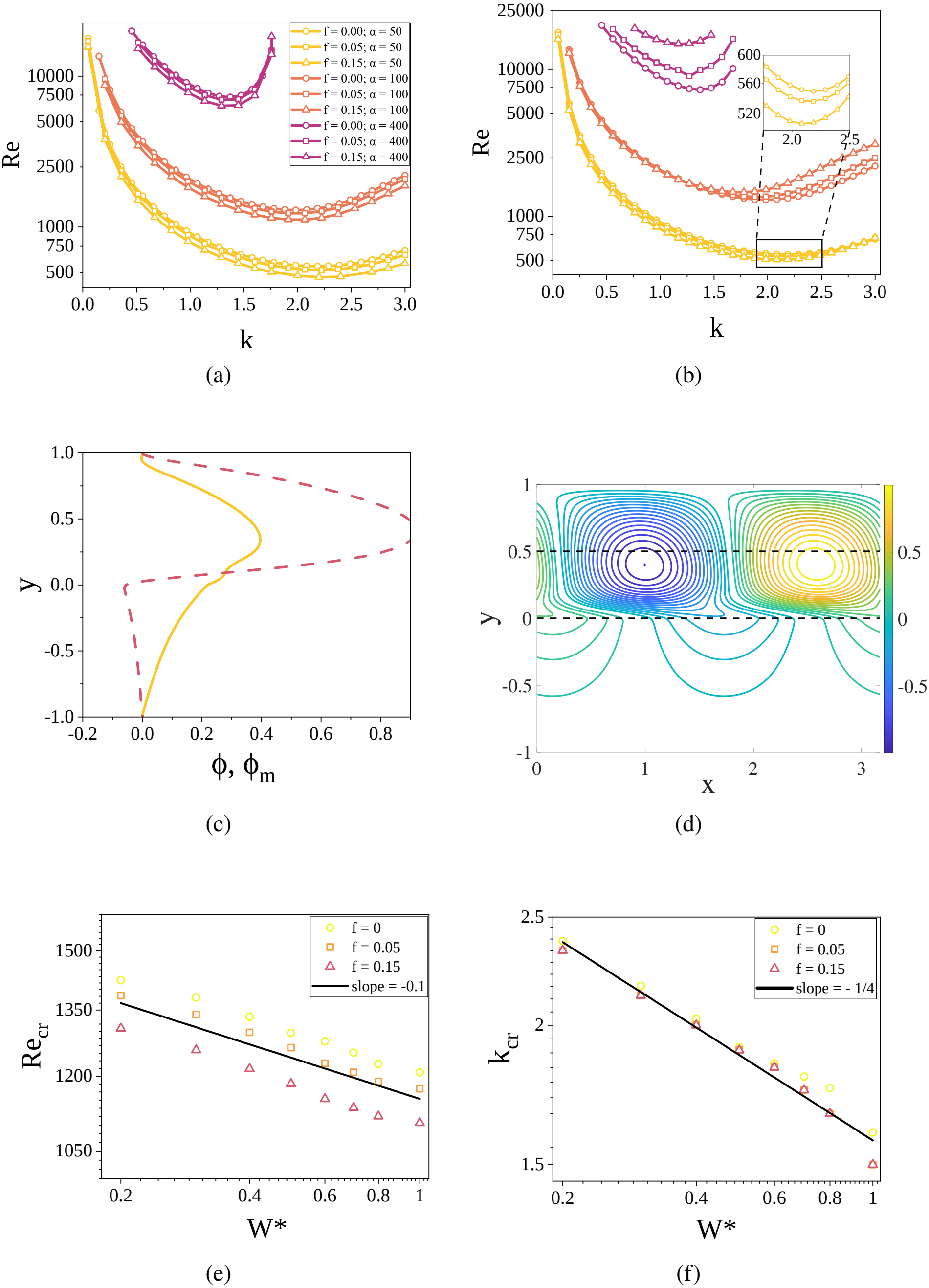

Figure 10: (a) The variation of the neutral curve in (k, Re) plane for $S = 1 \times 10^{-7}$ and $W^*$ = 0.5; (b) The variation of the neutral curve in (k, Re) plane for $S = 2.5 \times 10^{-4}$ and $W^*$ = 0.5; (c) The variation of normalized eigenfunction with coordinate y for $\alpha$=100, Re = 1295, k = 1.98, $f$ = 0.15, $S = 5 \times 10^{-5}$, $W^*$ =0.5; (d) The contour plot of associated stream function in (x, y) plane; (e) Variation of critical Reynolds number for different value of Couette component and particle mass fraction $f$ for $S = 5 \times 10^{-5}$, $\alpha$=100; (f) Variation of critical wave number for different value of Couette component and particle mass fraction $f$ for $S = 5 \times 10^{-5}$, $\alpha$=100. Other parameters: $\epsilon = 0.6$, $\tau = 0$, $\delta = 1$. Solid and dashed lines represent real and imaginary parts of the eigenfunction.

The variation of $Re_{cr}$ with respect to the Couette component, for a fixed particle relaxation time $S$ and varying mass fractions $f$, is presented in figure 10e. The results reveal a systematic decrease in $Re_{cr}$ with increasing upper-wall velocity across all particle mass fractions, following the approximate scaling $Re_{cr} \propto (W^*)^{-0.1}$. Increasing $W^*$ strengthens near-interface shear and increases momentum penetration into the porous layer, and enhancing interfacial disturbance amplification. This reduces the pressure driven Reynolds number required to excite the unstable branch, hence the observed monotonic decrease of $Re_{cr}$ with $W^*$. This behavior is consistent with the general destabilizing influence of increased permeability or penetration in porous-bounded channels (Samanta 2020). Moreover, this reduction becomes more pronounced at higher particle mass fractions, indicating that increased particle loading amplifies the destabilizing influence of the Couette component. This amplification is attributed to enhanced particle-fluid interactions and the associated momentum exchange in the presence of shear, which, together, lower the threshold for flow instability. A similar trend is observed for the critical wavenumber, $k_{cr}$, which decreases with increasing Couette component and follows the scaling $k_{cr} \propto (W^*)^{-1/4}$, as shown in figure 10f. The decrease of $k_{cr}$ with $W^*$ indicates that the most amplified disturbance becomes longer-wave as Couette forcing increases. Physically, stronger Couette motion redistributes the disturbance energy production, leading to a shift from short length scale structures concentrated near the interface toward larger scale shear structures extending across the fluid layer.This type of wavelength selection and mode shift under increasing Couette component is a known feature in porous-coupled Couette-Poiseuille systems (Chang *et al.* 2017), where increasing wall velocity can reorganize the neutral curve and change the dominant mode.

The impact of the Couette component on $Re_{cr}$ and critical wavenumber ($k_{cr}$) of the dominant fluid mode is systematically analyzed for varying particle relaxation times ($S$) at fixed particle mass fraction ($f$). The corresponding variations of $Re_{cr}$ and $k_{cr}$ as functions of the Couette component are shown in figure 11. For both particle-laden and particle-free cases, $Re_{cr}$ decreases as the Couette component increases as shown in figure 11a. The trend for the particle-free case is consistent with the observations reported by Samanta (2020). Interestingly, for all considered values $S$, the critical Reynolds number decreases in a nonlinear fashion with increasing Couette component $W^*$, following the approximate scaling laws $Re_{cr} \propto (W^*)^{-0.1}$. For intermediate $S$ (strong coupling), particle inertia adds effective dissipation and reduces the sensitivity of $Re_{cr}$ to W*(flattening trends). For very large $S$, coupling weakens and the particle phase approaches decoupling, so the trends revert toward the single-phase porous-wall behavior (consistent with the dusty-gas asymptotics). Similarly, $k_{cr}$ decreases as the Couette component increases, following the scaling law $k_{cr} \propto (W^*)^{-1/4}$ as shown in figure 11b.

To further elucidate the influence of the Couette component on flow stability, the variation in streamwise ($u$) and normal ($v$) perturbed velocity profiles corresponding to the most unstable modes is analyzed for both pure and particle-laden flows. These velocity profiles are presented in Figure 12 for a fixed permeability parameter ($\alpha$) at specific $k$ and $Re$ combinations. In the absence of the Couette component and porous layer, the streamwise velocity ($u$) in the particle-laden case exhibits a higher peak value than that of the pure fluid. Conversely, the normal velocity ($v$) in the pure fluid case exhibits a higher peak than in the particle-laden flow, as reported by Klinkenberg *et al.* (2011). This trend can be attributed to the dampening effect of particle inertia, which restricts the extent of vertical displacement in the particle-laden case.

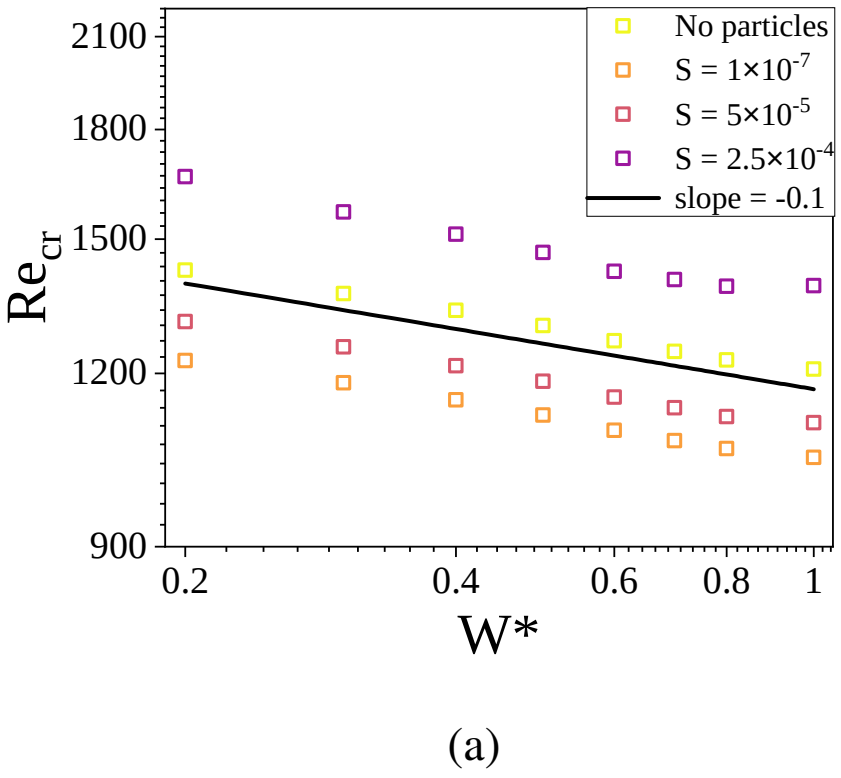

(a)

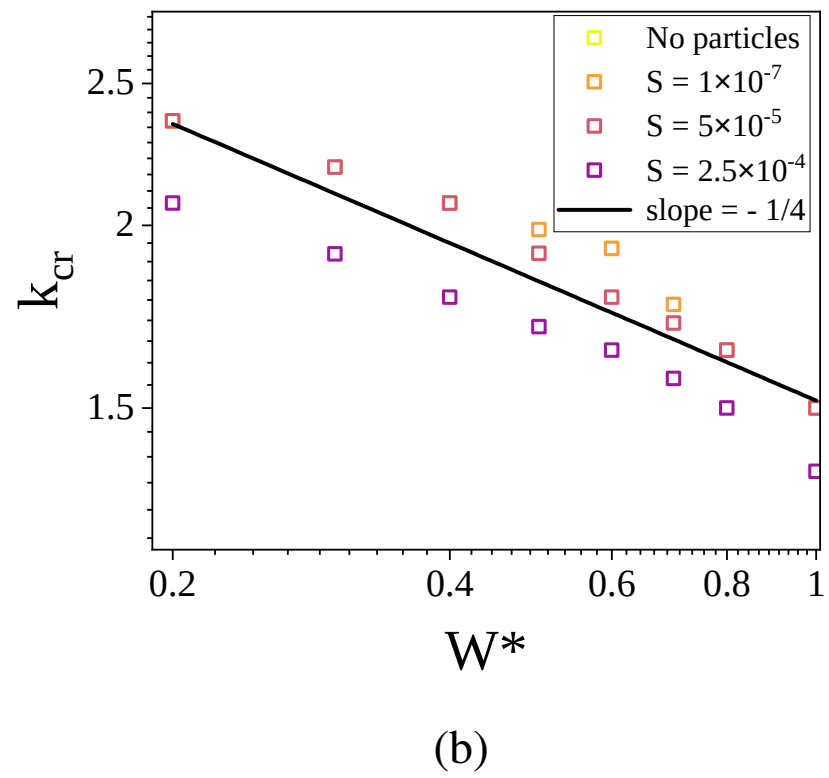

(b)

Figure 11: (a) Variation of the critical Reynolds number with Couette component for different particle relaxation times $S$; (b) Variation of the critical wavenumber with Couette component for different particle relaxation times $S$. The particle-free case ($f$=0) coincides with the curve for $S = 1 \times 10^{-7}$. Other parameters: $f = 0.15$, $\alpha$=100, $\epsilon = 0.6$, $\tau = 0$, $\delta = 1$.

Figure 12a illustrates the velocity profiles for $\alpha$=50, under conditions where the pure fluid case remains stable but the particle-laden flow (with $f = 0.15$) becomes unstable. Upon introducing a porous layer ($\alpha$=50), the maximum $u$-velocity becomes higher in the pure fluid case than in the particle-laden case, suggesting that the porous medium facilitates greater momentum diffusion in the absence of particles. Moreover, the symmetry of the streamwise velocity profile is disrupted, with the value of $u$ at the center of the fluid layer deviating from zero, unlike the symmetric profile observed in the absence of a porous layer. Figure 12b displays the corresponding velocity profiles for a higher permeability parameter, $\alpha$=400, under conditions where the pure fluid case is unstable while the particle-laden flow ($f$=0.15) is stabilized. In this scenario, the streamwise velocity profile in the particle-laden case more closely resembles that of the pure fluid, indicating that the particles' effect on the velocity field diminishes as permeability increases.

However, the disparity in the normal velocity ($v$) between the pure and particle-laden cases becomes more pronounced at higher permeability. The peak $v$-velocity is lower in the particle-laden case, suggesting that the porous layer exerts a stabilizing effect by dampening the vertical motion of the particles. Additionally, as the permeability increases, the streamwise velocity profile ($u$) becomes more symmetric, and the velocity at the center of the fluid layer approaches zero, indicating a return to the symmetric structure observed in purely fluid flows without a porous layer.

Figures 12c and 12d present the eigenfunctions of the streamwise ($u$) and wall-normal ($v$) velocity components for $\alpha = 50$ at representative values of $k$ and $Re$. In figure 12c, the chosen parameters correspond to a regime where the pure fluid case remains stable, while the particle-laden flow becomes unstable. Conversely, Figure 12d shows a regime where the pure fluid alone is unstable, but the addition of particles stabilizes the flow. In both cases, introducing the Couette component leads to a systematic difference between the pure and particle-laden profiles. The maximum $u$-velocity is consistently higher for the pure fluid compared to the suspension, indicating a suppression of streamwise motion by particle loading. Moreover, while the $u$-profile for the pure fluid remains nearly symmetric about the channel centerline, the particle-laden case exhibits a pronounced asymmetry, with the location of peak velocity shifted away from the mid-plane. This breaking of symmetry highlights the role of particle inertia in modifying the modal structure of the eigenfunctions under shear-Couette coupling. Overall, these results underscore the complex interplay between particle inertia, shear effects induced by the Couette component, and momentum transfer at the fluid-porous interface. The extent to which each factor influences the flow stability varies with both the relaxation time and the permeability parameter, as evidenced by the distinct trends in $Re_{cr}$, $k_{cr}$, and the velocity profiles.

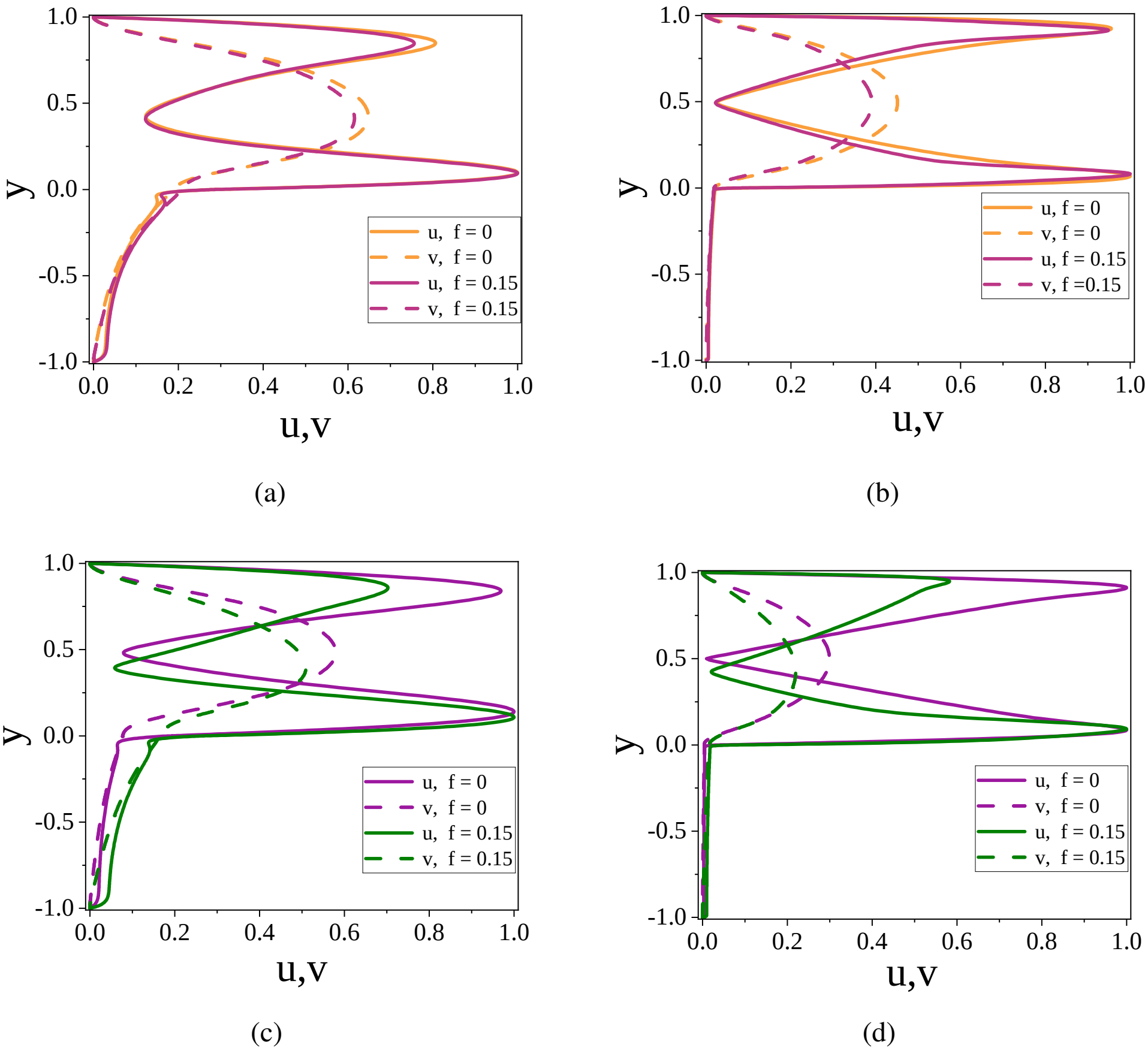

Figure 12: Variation of streamwise $u$ and normal $v$ perturbed velocity for flow with particles ($f$ = 0.15 and $S = 2.5 \times 10^{-4}$) and without particles. (a) $\alpha$=50, $W^*$ = 0, k = 2.7, Re = 690; (b) $\alpha$=400, $W^*$ = 0, k = 2, Re = 10000; (c) $\alpha$=50, $W^*$ = 0.5, k = 2.1, Re = 520; (d) $\alpha$=400, $W^*$ = 0.5, k = 1.2, Re = 10000. Other parameters: $\epsilon$ = 0.6, $\tau$ = 0, $\delta$ = 1.

### 3.4. *Limiting cases and comparison to subsystems*

The coupled suspension porous system recovers two well-defined limiting cases. In the unladen limit ($f = 0$), the formulation reduces to single-phase Couette-Poiseuille flow over a porous substrate. The neutral curves in figure 9 (for $\alpha = 50, 100, 400$) reproduce the established behavior of particle free porous wall flows: at low $\alpha$ (higher permeability), the instability exhibits bimodal structure due to competition between fluid and porous layer modes; as $\alpha$ increases, the curves become unimodal, dominated by the fluid layer mode as porous layer contributions weaken. This trend is consistent with previous studies of single-phase shear flows over porous beds (Chang *et al.* 2017; Samanta 2020; Hooshyar *et al.* 2022) and provides the baseline against which particle effects are assessed.

By contrast, for $\alpha \to \infty$, the porous layer approaches an effectively impermeable boundary, and the system reduces to classical particle-laden Couette Poiseuille flow. As $\alpha$ increases, porous associated eigenmodes are progressively damped (figure 5) and the neutral curves transition from bimodal to unimodal structure (figure 9). The resulting growth rates and critical Reynolds numbers approach those reported for particle-laden channel flows over rigid walls (Saffman 1962; Klinkenberg *et al.* 2011). In this limit, particle inertia retains its stabilizing influence for sufficiently large S, but the overall stability threshold is higher than in finite $\alpha$ cases owing to the removal of porous-induced destabilization.

## 4. Conclusions

This study presents a 3D linear stability analysis of particle-laden Couette-Poiseuille flow over a porous layer using a spectral collocation method. By adopting a two-domain approach, modeling the particle-laden fluid via Saffman's dusty-gas approximation with Stokes drag coupling, and the porous layer with the volume-averaged Navier-Stokes equations, the investigation systematically resolved the interplay between particle relaxation, porous permeability, and wall shear effects on modal stability.

The eigenspectrum analysis demonstrated that the canonical Y-shaped structure of plane Poiseuille flow over porous substrates is preserved in the presence of suspended particles. The coupled formulation generates additional disturbance branches associated with fluid-particle interaction in the presence of a permeable boundary. While these modes remain stable across the parameter space, their presence modifies the overall eigenspectrum and influences the dominant instability through altered coupling pathways, an effect not observed in either single-phase porous-wall flows.

A detailed parametric study revealed that particle effects are strongly mediated by porous permeability. At low permeability ($\alpha = 100$), particles with small relaxation times ($S = 5 \times 10^{-5}$) exert a destabilizing influence, lowering the critical Reynolds number. As permeability increases ($\alpha = 400$), this trend reverses, with stabilization dominating, highlighting a competition between porous-modified Tollmien-Schlichting modes and particle-fluid coupling. Across all tested relaxation times and mass fractions, the porous layer's destabilizing influence at low permeability parameter consistently outweighs the particle-induced stabilization. In classical particle-laden flows over impermeable walls, the influence of particle relaxation time $S$ and mass fraction $f$ on stability is generally non monotonic, with stabilization occurring over a finite range of $S$ where particle-fluid coupling is most effective. In the present system, however, the porous coating introduces a competing destabilizing mechanism through shear penetration into the porous layer. At sufficiently high permeability (low $\alpha$) and moderate $S$, particle addition can reduce $Re_{cr}$, in contrast to the impermeable-wall case. Classical stabilization is recovered only in the low-permeability limit. The results then indicate that particle inertia alone does not determine stability behavior; rather, its effect depends critically on boundary conditions, with porous walls reshaping the classical non monotonic dependence on relaxation time.

The inclusion of a Couette component introduced further destabilization. Although the eigenspectral topology remained qualitatively similar, wall modes underwent more pronounced variations, and perturbation growth rates increased with Couette strength. The neutral curves transitioned from bimodal to unimodal structures, and streamfunction profiles revealed enhanced penetration of disturbances into the porous layer. This mechanism explains why Couette forcing produced a robust destabilizing effect across all particle mass fractions and relaxation times, underscoring its dominant role in controlling modal amplification. In impermeable-wall Couette-Poiseuille flow, increasing the Couette component often stabilizes the system. Here, by contrast, $Re_{cr}$ decreases monotonically with $W^*$ across the explored parameter range.

Overall, this work demonstrates that the linear stability of particle-laden Couette-Poiseuille flow over porous walls is governed by a subtle balance between particle relaxation effects, porous permeability, and wall-driven shear. It reveals unanticipated results from coupling particle-laden flow with a porous boundary under combined Poiseuille-Couette forcing: (1) permeability-dependent reversal of particle stabilization, (2) emergence of new stable coupled modes that do not contribute to instability, and (3) monotonic destabilization by increasing Couette shear, overturning classical impermeable trends. These outcomes highlight the non-trivial interplay between Stokes drag dissipation, porous-layer penetration, and shear-driven interfacial modification. These findings extend classical stability analyses of dusty-gas suspensions and porous-wall flows, offering new physical insight into how multiphase interactions combined with Couette-driven shear and porous-wall effects alter the onset of instability. While the present dusty gas formulation captures the essential dynamics of dilute two-way coupling, future work could extend the analysis to denser regimes using two fluid models with particle stress closures or particle resolved simulations to incorporate collisions and finite size effects. It would also be of interest to examine nonlinear and transient dynamics beyond the linear framework considered here, including the interaction between modal instabilities and transport processes in porous suspension systems. Such efforts would be particularly relevant to engineering applications involving drag reduction, multiphase filtration, and biomedical flows through compliant or porous boundaries.

**Funding.** Supported by the National Science Foundation-CBET (Award No. 2230892).

**Declaration of interests.** The authors report no conflict of interest.

**Author ORCIDs.** Ananthapadmanabhan Ramesh, https://orcid.org/0009-0001-0459-100X;
Abbas Moradi Bilondii, https://orcid.org/0000-0003-3059-0470;
Mohammadreza Mahmoudian, https://orcid.org/0009-0004-7695-6487;
Parisa Mirbod, https://orcid.org/0000-0002-2627-1971.

## Appendix A. Eigenspectra of particle-laden channel flows with and without porous layers

In this appendix, we discuss the fundamental characteristics of the eigenspectrum for particle-laden Poiseuille flow with and without a porous layer. The focus is on the emergence and evolution of the most unstable mode, as well as the distribution and trajectory of other discrete, stable modes, which are systematically presented through eigenspectral analysis.

### A.1. *Particle-laden flows spectra in a plane Poiseuille flow (PF)*

In this appendix, we provide a detailed analysis of the eigenspectral characteristics of particle-laden Poiseuille flow in a channel of width $2L$, comparing them with those of classical plane Poiseuille flow. The influence of suspended particles is systematically investigated by varying the particle relaxation time ($S$) and mass fraction ($f$). In the absence of particles ($f$=0), flow reverts to that of the classical plane Poiseuille flow. This baseline case is depicted in figure 13a, where the eigenspectrum at the critical Reynolds number $Re = 5772.22$ and streamwise wavenumber $k = 1.02$ exhibits a distinctive Y-shaped structure, characterized by the distribution of eigenvalues across three clearly identifiable branches. This branching pattern arises due to differences in the spatial characteristics of the eigenfunctions, particularly their behavior near the channel center and walls. The upper right branch of eigenvalues corresponds to center modes, whose eigenfunctions exhibit dominant variations near the channel center. The upper-left branch consists of wall modes, associated with eigenfunctions that exhibit significant variation near the channel walls. Lastly, the lower branch consists of damped modes that decay over time and do not contribute to instability (Schmid *et al.* 2002).

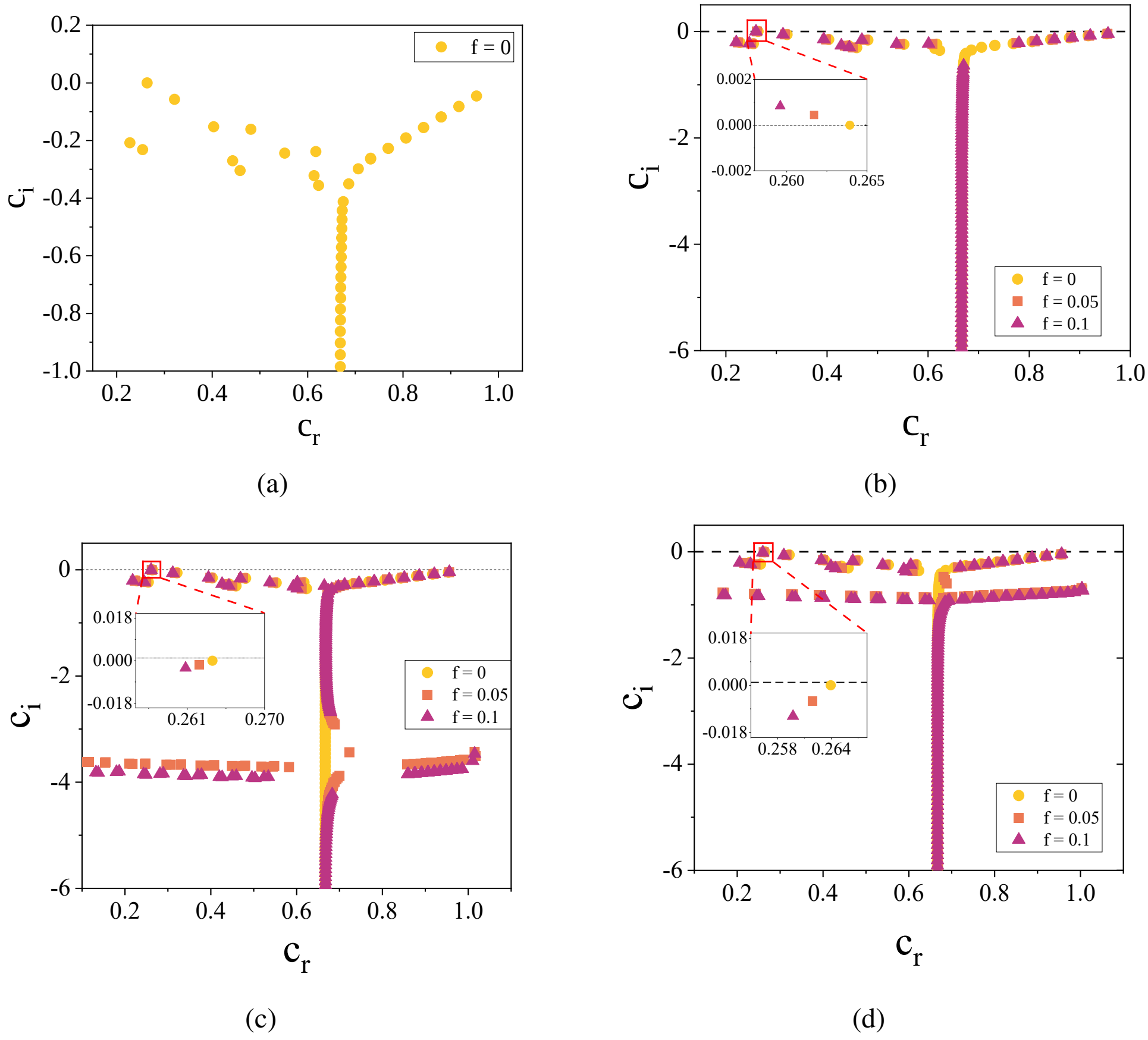

Figure 13: Eigenspectra variation for different particle parameters (relaxation time S and mass fraction f) in a channel of width 2L without Couette component and porous layer. (a) Poiseuille flow without particles (b) $S = 1 \times 10^{-7}$; (c) $S = 5 \times 10^{-5}$; (d) $S = 2.5 \times 10^{-4}$. Other parameters: $Re = 5772.22$, $k = 1.02$, $\epsilon$=0.6, $\tau = 0$, $\delta$=1.

At these critical conditions, the growth rate ($c_i$) of the most unstable mode is close to zero, indicating marginal stability, while all other eigenvalues have negative $c_i$, confirming a stable flow regime. For $Re > 5772$, the wall mode becomes unstable, corresponding to the onset of the TS instability. This behavior serves as a reference for analyzing the effects of particle addition. The impact of increasing the particle mass fraction is shown in figures 13b-13d, where the eigenspectrum is presented for $f$ ranging from 0 to 0.15 for different values of $S$. The objective here is to assess how variations in $f$ affect the least stable mode and its interaction with other modes at different relaxation times. As the mass fraction increases, the distribution of eigenvalues and the growth rate of the least stable mode, particularly the TS mode, change.

Next, the influence of particle relaxation time is examined. Figure 13b shows the eigenspectrum for a small relaxation time ($S = 1 \times 10^{-7}$). In this scenario, the characteristic Y-shaped structure of the pure fluid case ($f = 0$) is largely retained, with only minor shifts in the eigenvalue distribution. However, as the mass fraction $f$ increases, these shifts become more pronounced, leading to a noticeable increase in the TS mode's growth rate and further indicating destabilization.

The effect of higher relaxation times is depicted in figures 13c and 13d, corresponding to $S = 5 \times 10^{-5}$ and $S = 2.5 \times 10^{-4}$, respectively. In these cases, the presence of particles stabilizes the flow, as evidenced by the reduction in the growth rate of the TS mode. The most unstable mode becomes progressively damped as the relaxation time increases, suggesting that particles with higher $S$ act as stabilizers, mitigating the destabilizing effects observed at lower $S$. These results indicate that while increasing the particle mass fraction generally amplifies the growth rate of the TS mode, increasing the relaxation time counteracts this effect, thereby stabilizing the flow. This dual influence of particle parameters on flow stability provides valuable insights into the interplay between particle dynamics and flow stability in particle-laden Poiseuille flows.

### A.2. *Eigenspectra of particle-laden Poiseuille flow overlying a porous layer with $\alpha$=50*

In this section, we investigate the effects of particle addition on the eigenspectra for a channel flow with a permeability parameter of $\alpha$=50. This value represents a relatively high channel permeability compared to higher values such as $\alpha$= 100 and $\alpha$= 400. By analyzing the response of the eigenspectrum to varying particle relaxation times ($S$) and mass fractions ($f$), we aim to elucidate the destabilizing mechanisms in particle-laden flows over a porous medium. Figures 14a-14c show the evolution of the eigenspectrum for $\alpha$=50 when particles are introduced with different $S$ and $f$. At a small relaxation time of $S = 1 \times 10^{-7}$, as shown in figure 14a, the porous-modified TS mode, which is initially stable in the absence of particles ($f$=0), undergoes significant amplification as the mass fraction increases. This transition leads to instability, with the porous-modified TS mode becoming unstable ($c_i > 0$). Moreover, while Modes 2 and 3 exhibit increased growth rates with increasing $f$, they remain below the neutral stability threshold ($c_i = 0$), indicating that these modes remain stable despite particle addition. Mode 4, however, shows minimal sensitivity to particle effects, maintaining its initial stability profile regardless of changes in $f$. Notably, the growth rate of the porous-modified TS mode is higher for $\alpha = 50$ than for higher permeability parameters such as $\alpha = 100$ and $\alpha = 400$, indicating that at lower permeability parameter, the channel is more susceptible to particle-induced destabilization.

The corresponding eigenspectrum for particles with an intermediate relaxation time of $S = 5 \times 10^{-5}$ is presented in figure 14b. As in the case of a smaller $S$, the porous-modified TS mode exhibits a noticeable increase in growth rate as $f$ increases. The increased relaxation time allows particles to respond more effectively to fluid perturbations, amplifying the instability in the porous-modified TS mode. The remaining modes (Modes 2, 3, and 4) exhibit minor variations in growth rates, but none transition to instability. Figure 14c demonstrates the eigenspectrum for particles with a larger relaxation time of $S = 2.5 \times 10^{-4}$. At this relaxation time, particles can maintain momentum over longer time scales, potentially enhancing destabilizing interactions within the porous layer. The results indicate that the porous-modified TS mode, initially stable in the absence of particles, becomes progressively more unstable with increasing $f$. This trend is consistent with the observations for smaller $S$, but the destabilizing effect is more substantial at higher $S$. Moreover, at $S = 2.5 \times 10^{-4}$, the growth rates of Modes 2 and 3 also show slight amplification as $f$ increases, though they remain below the neutral stability threshold. Mode 4 remains unaffected by particle addition, suggesting that particles primarily affect the TS mode and its adjacent modes.

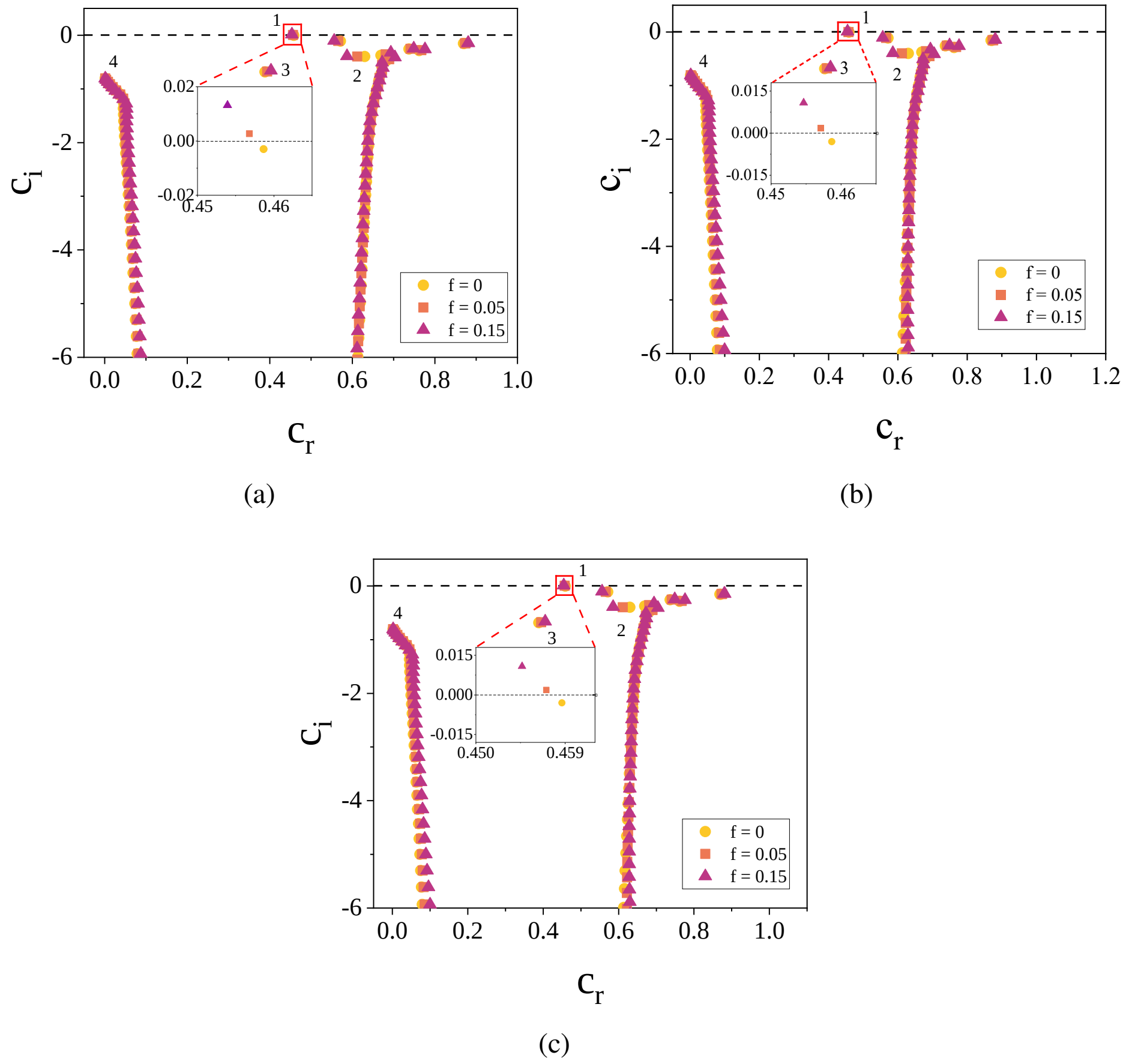

Figure 14: Eigenspectra variation for different particle parameters (relaxation time S and mass fraction f) for flow with porous layer of $\alpha$ =50 and without Couette component. (a) $S = 1 \times 10^{-7}$; (b) $S = 5 \times 10^{-5}$; (c) $S = 2.5 \times 10^{-4}$. Other parameters: $Re$ = 680, $k$ = 2.8, $\epsilon$=0.6, $\tau$=0, $\delta$=1.

## Appendix B. Variation of eigenfunctions and streamfunction fields with the permeability parameter

The addition of particles to a channel without a porous layer typically results in flow stabilization, as demonstrated by Klinkenberg *et al.* (2011), except in cases where the particle relaxation time is very small, such as $S = 1\times10^{-7}$. However, the introduction of a porous layer significantly alters this behavior, especially at lower values of the permeability parameter $\alpha$. To better understand the influence of porous media on flow stability, we analyze the eigenfunctions and corresponding stream functions by varying $\alpha$.

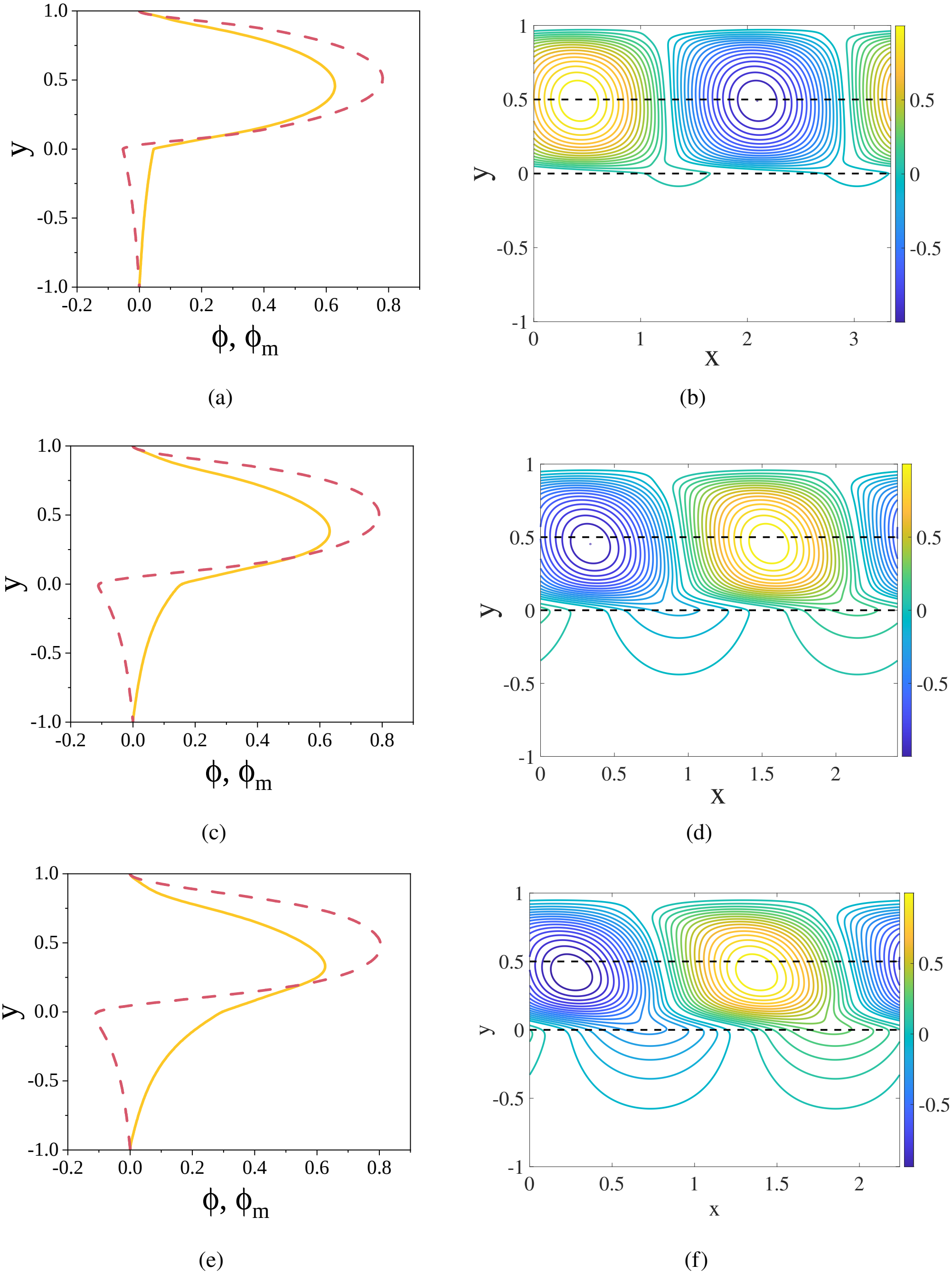

Figure 15: Normalized eigenfunctions ($\phi$, $\phi_m$) and corresponding streamfunction contours in the $(x, y)$ plane for different values of the permeability parameter $\alpha$. (a, b) $\alpha$ = 400, $Re_{cr}$ = 14496, $k_{cr}$ = 1.88; (c, d) $\alpha$ = 100, $Re_{cr}$ = 1490.8, $k_{cr}$ = 2.6; (e, f) $\alpha$ = 50, $Re_{cr}$ = 608, $k_{cr}$ = 2.8. Other parameters: $f$ = 0.15, $S = 5 \times 10^{-5}$, $\delta$ = 1, $\epsilon$ = 0.6, $W^*$ = 0, $\tau$ = 0. Solid and dashed lines represent real and imaginary parts of the eigenfunction.

In this study, the absolute minimum of the neutral curve, where instability first emerges, is observed within the fluid mode. To systematically investigate, the permeability parameter $\alpha$ is varied, and for each case, the Reynolds number and wavenumber are selected from the corresponding unstable region in the fluid mode for the single-phase flow (i.e., without particles). Figure 15a illustrates the eigenfunction profile for a case where the fluid mode is dominant at $\alpha$=400. At the selected Reynolds number and wavenumber, no flow reversal is observed in the central region of the fluid layer. This is consistent with the findings reported by Chang *et al.* (2006), Hill & Straughan (2008*b*), Liu *et al.* (2008*b*), and Hill & Straughan (2009). However, despite the absence of flow reversal, a pronounced variation in the eigenfunction is evident in the fluid domain, particularly at the center of the channel. The corresponding stream function in figure 15b reveals the formation of vortices in the central fluid layer, serving as the primary source of instability. Although the vortices are primarily confined to the fluid region, a portion of the vortex structure penetrates the porous medium due to momentum diffusion at the fluid-porous interface, as discussed by Samanta (2020).

As the permeability parameter $\alpha$ decreases, the flow's penetration into the porous layer increases, leading to significant changes in the eigenfunction's structure near the fluid-porous interface. This effect is clearly observed in figures 15d and 15f illustrate how vortex motion propagates deeper into the porous medium, driven by enhanced momentum diffusion from the fluid layer. This enhanced vortex penetration at lower $\alpha$ has important implications for the stability of particle-laden flows. In particular, the stabilizing effect of particles with relatively long relaxation times (e.g., $S = 5 \times 10^{-5}$ and $S = 2.5 \times 10^{-4}$) is significantly diminished in the presence of a high-permeability porous layer (low $\alpha$). As the flow penetrates further into the porous medium, the particle-laden structure interacts more effectively with the porous substrate, thereby reducing the stabilizing influence of particle inertia. This critical behavior highlights the critical role of permeability in modulating the stability characteristics of particle-laden flows through porous channels.

## Appendix C. Applicability of Squire's theorem for particle-laden Couette–Poiseuille flow over a porous medium

To examine the dimensionality of the least stable modes, we consider three-dimensional perturbations of the form $\tilde{\xi}(x, y, z, t) = \hat{\xi}(y)e^{i(k_x x + k_z z - ct)}$, where $\hat{\xi}(y) = (\hat{u}, \hat{v}, \hat{p}, \hat{n}, \hat{u_m}, \hat{v_m}, \hat{p_m})$ in linearized governing equations 2.15 to 2.32 yields

$$\mathrm{i}k_x\hat{u} + D\hat{v} + ik_z\hat{w} = 0, \tag{C 1}$$

$$\left(Uik_x - \frac{1}{Re}(D^2 - (k_x^2 + k_z^2)) + \frac{f}{SRe}\right)\hat{u} + U'\hat{v} - \frac{f}{SRe}\hat{u}_p + \mathrm{i}k_x\hat{p} = ik_x c\hat{u}, \tag{C 2}$$

$$\left(Uik_x - \frac{1}{Re}(D^2 - (k_x^2 + k_z^2)) + \frac{f}{SRe}\right)\hat{v} - \frac{f}{SRe}\hat{v}_p + D\hat{p} = ik_x c\hat{v}. \tag{C 3}$$

$$\left(Uik_x - \frac{1}{Re}(D^2 - (k_x^2 + k_z^2)) + \frac{f}{SRe}\right)\hat{w} - \frac{f}{SRe}\hat{w}_p + \mathrm{i}k_z\hat{p} = ik_x c\hat{w}, \tag{C 4}$$

For the particle phase, the governing equations can be reformulated as

$$\mathrm{i}k_x\hat{u}_p + D\hat{v}_p + U\mathrm{i}k_x\hat{n} + \mathrm{i}k_z\hat{w}_p = ikc\hat{n}, \tag{C 5}$$

$$-\frac{1}{SRe}\hat{u}+\left(Uik_x+\frac{1}{SRe}\right)\hat{u}_p+U'\hat{v}_p=ik_xc\hat{u}_p, \tag{C 6}$$

$$-\frac{1}{SRe}\hat{v}+\left(Uik_x+\frac{1}{SRe}\right)\hat{v}_p=ik_xc\hat{v}_p. \tag{C 7}$$

$$-\frac{1}{SRe}\hat{w}+\left(Uik_x+\frac{1}{SRe}\right)\hat{v}_p=ik_xc\hat{w}_p. \tag{C 8}$$

Similarly, for the porous layer, the equations take the form of

$$\mathrm{i}k_x\hat{u}_m+D\hat{v}_m+ik_z\hat{w}_m=0, \tag{C 9}$$

$$\left(\frac{\epsilon\alpha^2}{Re}-\frac{1}{Re}\left(D_m^2-(k_x^2+k_z^2)\right)\right)\hat{u}_m+\epsilon\mathrm{i}k_x\hat{p}_m=ik_xc\hat{u}_m, \tag{C 10}$$

$$\left(\frac{\epsilon\alpha^2}{Re}-\frac{1}{Re}\left(D_m^2-(k_x^2+k_z^2)\right)\right)\hat{v}_m+\epsilon D\hat{p}_m=ik_xc\hat{v}_m. \tag{C 11}$$

$$\left(\frac{\epsilon\alpha^2}{Re}-\frac{1}{Re}\left(D_m^2-(k_x^2+k_z^2)\right)\right)\hat{w}_m+\epsilon\mathrm{i}k_z\hat{p}_m=ik_xc\hat{w}_m. \tag{C 12}$$

In the above equations C 1 - C 12, different parameters are defined as $D=\frac{\partial}{\partial y}$, $D^2=\frac{\partial^2}{\partial y^2}$, $D_m=\frac{\partial}{\partial y_m}$, $D_m^2=\frac{\partial^2}{\partial y_m^2}$ and $U'=dU/dy$. By applying perturbation to the linearized boundary conditions (equations 2.27 - 2.32 ) yields:

$$\hat{u}=\hat{v}=\hat{w}=0 \qquad \text{at} \qquad y=1, \tag{C 13}$$

$$\hat{u}=\hat{u}_m \quad , \quad \hat{v}=\hat{v}_m \quad , \quad \hat{w}=\hat{w}_m \qquad \text{at} \qquad y=0, \tag{C 14}$$

$$\hat{p}_m-\frac{2}{\epsilon Re}D\hat{v}_m=\hat{p}-\frac{2}{Re}D\hat{v} \qquad \text{at} \qquad y=0, \tag{C 15}$$

$$\frac{1}{\epsilon}D\hat{u}_m-D\hat{u}=\tau\alpha\hat{u}_m \qquad \text{at} \qquad y=0, \tag{C 16}$$

$$\frac{1}{\epsilon}D\hat{w}_m-D\hat{w}=\tau\alpha\hat{w}_m \qquad \text{at} \qquad y=0, \tag{C 17}$$

$$\hat{u}_m=\hat{v}_m=\hat{w}_m=0 \qquad \text{at} \qquad y=-\delta. \tag{C 18}$$

It is observed that the density field ($\hat{n}$) becomes a passive scalar and is independent of all equations except equation C 5. Therefore, the evolution of $\hat{n}$ can be neglected (Saffman 1962; Sozza *et al.* 2022). By means of a Squire's transformation, $k_x\hat{u}+k_z\hat{w}=\tilde{k}\tilde{u}$, $\hat{v}=\tilde{v}$, $\tilde{p}=k\hat{p}/k_x$, $\tilde{k}=\sqrt{k_x^2+k_z^2}$, $k_xRe=\tilde{k}\tilde{R}e$, $k_x\hat{u}_p+k_z\hat{w}_p=\tilde{k}\tilde{u}_p$, $\hat{v}_p=\tilde{v}_p$, $k_x\hat{u}_m+k_z\hat{w}_m=\tilde{k}\tilde{u}_m$, $\tilde{p}_m=k\hat{p}_m/k_x$, $\hat{v}_m=\tilde{v}_m$, the three dimensional disturbance problem is mapped onto an equivalent two dimensional system with streamwise wavenumber $\tilde{k}$ and Reynolds number $\tilde{R}e$. The governing equations for the fluid, particle, and porous phases retain the same form as the two-dimensional stability equations, given by

$$\mathrm{i}\tilde{k}\tilde{u}+D\tilde{v}=0, \tag{C 19}$$

$$\left(Ui\tilde{k}-\frac{1}{\tilde{R}e}(D^2-\tilde{k}^2)+\frac{f}{S\tilde{R}e}\right)\tilde{u}+U'\tilde{v}-\frac{f}{SRe}\tilde{u}_p+\mathrm{i}\tilde{k}\tilde{p}=i\tilde{k}c\tilde{u}, \tag{C 20}$$

$$\left(Ui\tilde{k} - \frac{1}{\tilde{Re}}(D^2 - \tilde{k}^2) + \frac{f}{S\tilde{Re}}\right)\tilde{v} - \frac{f}{S\tilde{Re}}\tilde{v}_p + D\tilde{p} = i\tilde{k}c\tilde{v}. \tag{C 21}$$

$$-\frac{1}{S\tilde{Re}}\tilde{u} + \left(Ui\tilde{k} + \frac{1}{S\tilde{Re}}\right)\tilde{u}_p + U'\tilde{v}_p = i\tilde{k}c\tilde{u}_p, \tag{C 22}$$

$$-\frac{1}{S\tilde{Re}}\tilde{v} + \left(Ui\tilde{k} + \frac{1}{S\tilde{Re}}\right)\tilde{v}_p = i\tilde{k}c\tilde{v}_p. \tag{C 23}$$

$$\mathrm{i}\tilde{k}\tilde{u}_m + D\tilde{v}_m = 0, \tag{C 24}$$

$$\left(\frac{\epsilon\alpha^2}{\tilde{Re}} - \frac{1}{\tilde{Re}}\left(D_m^2 - \tilde{k}^2\right)\right)\tilde{u}_m + \epsilon\mathrm{i}\tilde{k}\tilde{p}_m = i\tilde{k}c\tilde{u}_m, \tag{C 25}$$

$$\left(\frac{\epsilon\alpha^2}{\tilde{Re}} - \frac{1}{\tilde{Re}}\left(D_m^2 - \tilde{k}^2\right)\right)\tilde{v}_m + \epsilon D\tilde{p}_m = i\tilde{k}c\tilde{v}_m. \tag{C 26}$$

$$\tilde{u} = \tilde{v} = 0 \qquad \text{at} \qquad y = 1, \tag{C 27}$$

$$\tilde{u} = \tilde{u}_m \quad , \quad \tilde{v} = \tilde{v}_m \qquad \text{at} \qquad y = 0, \tag{C 28}$$

$$\tilde{p}_m - \frac{2}{\epsilon\tilde{Re}}D\tilde{v}_m = \tilde{p} - \frac{2}{\tilde{Re}}D\tilde{v} \qquad \text{at} \qquad y = 0, \tag{C 29}$$

$$\frac{1}{\epsilon}D\tilde{u}_m - D\tilde{u} = \tau\alpha\tilde{u}_m \qquad \text{at} \qquad y = 0, \tag{C 30}$$

$$\tilde{u}_m = \tilde{v}_m = 0 \qquad \text{at} \qquad y = -\delta. \tag{C 31}$$

Under this mapping, the three-dimensional system reduces exactly to the two-dimensional disturbance system with streamwise wavenumber $\tilde{k}$ and Reynolds number $\tilde{Re}$. The transformed equations retain the same structure as the two-dimensional stability problem presented in the main text. The boundary conditions are invariant under this transformation, and therefore, the entire eigenvalue problem maps consistently onto its two-dimensional counterpart. Since the transformed Reynolds number satisfies $\tilde{Re} \leqslant Re$ for a given $k_x$, the minimum critical Reynolds number is attained for purely two-dimensional disturbances ($k_z = 0$). Consequently, the least stable modes of the particle-laden Couette–Poiseuille flow over a porous medium are two-dimensional, and the linear stability analysis may be confined to two-dimensional perturbations without loss of generality.